\documentclass[12pt]{article} 
\usepackage{amssymb}
\usepackage{amsmath} 
\usepackage{cprotect}
\usepackage{cite}
\usepackage{axodraw}
\usepackage{youngtab}
\usepackage{hyperref}
\input xy
\xyoption{all}
\usepackage{enumitem}

\def\sqr#1#2{{\vcenter{\vbox{\hrule height.#2pt
            \hbox{\vrule width.#2pt height#1pt \kern#1pt
                  \vrule width.#2pt}\hrule height.#2pt}}}}

\def\sqra#1#2#3{{\vcenter{\vbox{\hrule height.#2pt
            \hbox{\vrule width.#2pt height#1pt \kern5pt 
#3
                  \vrule width.#2pt}\hrule height.#2pt}}}}

\numberwithin{equation}{section}

\numberwithin{table}{section}

\begin{document} 

\begin{center}

{\large\bf Decomposition in Chern-Simons theories in three dimensions}

\vspace*{0.2in}

Tony Pantev$^1$, Eric Sharpe$^2$

\begin{tabular}{cc}
{\begin{tabular}{l}
$^1$ Department of Mathematics\\
David Rittenhouse Lab.\\
209 South 33rd Street\\
Philadelphia, PA  19104-6395 \end{tabular}}
&
{\begin{tabular}{l}
$^2$ Department of Physics MC 0435\\
850 West Campus Drive\\
Virginia Tech\\
Blacksburg, VA  24061 \end{tabular}}
\end{tabular} 

{\tt tpantev@math.upenn.edu},
{\tt ersharpe@vt.edu}

\end{center}

In this paper we discuss decomposition in the context of three-dimensional
Chern-Simons theories.  Specifically, we argue that a Chern-Simons theory
with a gauged 
noneffectively-acting one-form symmetry is
equivalent to a disjoint union of Chern-Simons theories, with 
discrete theta angles coupling to the image under a Bockstein homomorphism
of a canonical degree-two characteristic class.  On three-manifolds with
boundary, we show that the bulk discrete theta angles 
(coupling to bundle characteristic classes)
are mapped to choices of discrete torsion in boundary orbifolds.
We use this to verify that the bulk three-dimensional Chern-Simons decomposition
reduces on the boundary
to known decompositions of two-dimensional (WZW) orbifolds, 
providing a strong consistency test of our proposal.

\begin{flushleft}
June 2022
\end{flushleft}

\newpage

\tableofcontents

\newpage

\section{Introduction}

Decomposition is the observation that some local quantum field theories
are equivalent to disjoint unions of other local quantum field theories,
essentially a counterexample to old lore linking locality and cluster
decomposition.  It was first\footnote{
For purposes of historical
language translation, before the term `one-form symmetry'
was coined, theories with one-form symmetries were sometimes called
`gerby' theories, in reference to the fact that a gerbe is a fiber bundle
whose fibers are higher groups.
} observed in
\cite{Hellerman:2006zs} in two-dimensional gauge theories and
orbifolds with trivially-acting subgroups (nonminimally-charged
matter) \cite{Pantev:2005rh,Pantev:2005wj,Pantev:2005zs},
and since then has been developed in many other references,
see e.g.~\cite{Pantev:2022kpl,Robbins:2020msp,Robbins:2021ibx,Robbins:2021lry,Robbins:2021xce,Yu:2021zmu,Tanizaki:2019rbk,Cherman:2020cvw,Cherman:2021nox,Sharpe:2014tca,Eager:2020rra,Anderson:2013sia,Sharpe:2019ddn,Komargodski:2020mxz,ajt1,ajt2,ajt3,t1,gt1,xt1,Caldararu:2010ljp,Hellerman:2010fv,Nguyen:2021yld,Nguyen:2021naa,Honda:2021ovk,Huang:2021zvu}
and e.g.~\cite{Sharpe:2006vd,Sharpe:2010zz,Sharpe:2010iv,Sharpe:2019yag,Sharpe:2022ene} for reviews.

Decomposition is not limited to two dimensions, and indeed
four-dimensional versions of decomposition have been
described in \cite{Tanizaki:2019rbk,Cherman:2020cvw}.
The common thread linking these different examples involves
what is now called a higher-form symmetry:  a quantum field theory
in $d$ spacetime dimensions decomposes if it has a global
$(d-1)$-form symmetry (possibly realized noninvertibly)
\cite{Tanizaki:2019rbk,Cherman:2020cvw}.

In this paper, following up \cite{Pantev:2022kpl}, we turn to
decomposition in three-dimensional Chern-Simons theories with
gauged noneffectively-acting one-form symmetries.
Briefly, we find that
\begin{equation} \label{eq:main}
\left[ \mbox{Chern-Simons}(H) / BA \right] \: = \:
\coprod_{\theta \in \hat{K}} \mbox{Chern-Simons}(G)_{\theta},
\end{equation}
where $G = H / (A/K)$,
$K \subset A$ defines the trivially-acting subgroup,
and
$\theta$ indicates a discrete theta angle coupling to an
appropriate characteristic class of $G$ bundles,
On the boundary, this reduces to decomposition in noneffectively-acting
orbifolds of two-dimensional WZW models.  A key role is played by the
fact that the bulk discrete theta angles (coupling to bundle characteristic
classes) become discrete torsion on the boundary,
a result we explain in detail.
The fact that the bulk decomposition correctly
implies a known decomposition of the
two-dimensional boundary theory
provides a strong consistency
check on our proposal.

In two dimensions, decomposition has had a variety of applications,
for example in
giving nonperturbative constructions of geometries in phases of
some gauged linear sigma models (GLSMs) 
\cite{Caldararu:2010ljp,Hori:2011pd,Addington:2012zv,Sharpe:2012ji,Halverson:2013eua,Ballard:2013fxa,Sharpe:2013bwa,Hori:2013gga,Hori:2016txh,Wong:2017cqs,Kapustka:2017jyt,Parsian:2018fhm,Chen:2018qww,Guo:2021aqj}, 
in Gromov-Witten theory \cite{ajt1,ajt2,ajt3,t1,gt1,xt1}, 
in computing elliptic genera to
check claims about IR limits of pure supersymmetric gauge
theories \cite{Eager:2020rra}, and recently in understanding 
Wang-Wen-Witten anomaly resolution 
\cite{Wang:2017loc,Robbins:2021ibx,Robbins:2021lry,Robbins:2021xce}.

Chern-Simons theories are the starting point for many physics questions,
and so we anticipate that the results of this paper 
should have a variety of applications.
For example, as is well-known, three-dimensional AdS gravity can be
understood as a Chern-Simons theory \cite{Witten:1988hc},
making Chern-Simons theories a natural playground for addressing
questions in three-dimensional gravity, an approach used in 
e.g.~\cite{Benini:2022hzx} to address
Marolf-Maxfield factorization questions \cite{Marolf:2020xie}.
We anticipate that this work may have analogous uses.

Similarly, one of the original applications of two-dimensional decomposition 
was to understand phases of certain gauged linear sigma models,
where decomposition was used locally (ala Born-Oppenheimer)
to understand IR limits of certain theories as nonperturbatively-realized
branched covers of 
spaces \cite{Caldararu:2010ljp}.
We expect that similar ideas could be used to understand the IR
limits of certain Chern-Simons-matter theories.

We begin in section~\ref{sect:review-wzw} with a review of decomposition in
two-dimensional WZW orbifolds, which not only serves as a review of
decomposition, but also describes the decomposition pertinent to
boundaries in the three-dimensional Chern-Simons theories we discuss.

In section~\ref{sect:decomp} we describe the primary proposal of this
paper, namely decomposition in Chern-Simons theories with gauged
one-form symmetry groups, which takes the form~(\ref{eq:main}).
All Chern-Simons theories are assumed to have levels such that the theories
exist on the three manifolds over which they are defined.
We describe how this bulk decomposition maps to boundary WZW models,
and reproduces standard results on decomposition
in two-dimensional noneffective orbifolds, which serves as a strong consistency
test of our claims.  We also observe that in all these examples,
the boundary discrete theta angles (choices of discrete torsion in boundary
WZW models) are trivial, which is often reflected in the bulk discrete
theta angles.

In section~\ref{sect:spectra} we discuss the spectra of these theories.
We begin with an explanation and review of monopole operators, local
operators (analogues of twist fields in two-dimensional orbifolds) which
can be used to construct projection operators.  We then discuss
line operators.  
When gauging ordinary one-form symmetries, the standard technology of anyon
condensation can be used to describe the line operators.
However, to describe
noneffectively-acting one-form symmetries (in which a subgroup acts
trivially), as relevant for this paper,
requires a minor extension, which we propose and utilize.

In section~\ref{sect:exs} we walk through the details of bulk and
boundary decomposition, spectrum computations, and consistency tests
such as level-rank duality in a variety of concrete examples.

Finally in section~\ref{sect:boundary-g-g} we briefly discuss
the related case of boundary $G/G$ models.
These two-dimensional theories decompose, and we briefly discuss
their corresponding bulk theories.

In appendix~\ref{app:lineops} we summarize some results on line operators
that are used in the main text.  In appendix~\ref{app:crossed-module} we
give a brief overview of crossed modules, to make this paper self-contained,
as they are used in the description of three-dimensional decomposition.
In appendix~\ref{sect:gauging-bk} we describe gauging 
effectively-acting one-form symmetries without appealing to line operators.

\section{Warm-up: Decomposition in WZW orbifolds}
\label{sect:review-wzw}

As a warm-up exercise, let us briefly review decomposition in two dimensions,
and apply it towards
orbifolds of WZW models.

Consider an orbifold $[X/\Gamma]$ where a central subgroup $K \subset
\Gamma$ acts trivially on $X$.  As has been discussed previously
(see e.g.~\cite{Hellerman:2006zs}),
for an ordinary (orientation-preserving) orbifold,
\begin{equation}  
{\rm QFT}\left( [X/\Gamma] \right) \: = \:
\coprod_{\theta \in \hat{K}} 
{\rm QFT}\left( [X/G]_{\theta(\omega)} \right),
\end{equation}
where $\theta(\omega)$ is a choice of discrete torsion,
given as the image of the extension class
$[\omega] \in H^2(G,K)$ corresponding to
\begin{equation}
1 \: \longrightarrow \: K \: \longrightarrow \: \Gamma \:
\longrightarrow \: G \: \longrightarrow \: 1
\end{equation}
under the map $\theta: K \rightarrow U(1)$,
yielding $\theta(\omega) \in H^2(K,U(1))$.

Consider a $\Gamma$ orbifold of a WZW model for a group $H$,
with $K \subset \Gamma$ acting trivially, and $G = \Gamma/K$ a subset
of the center of $H$, acting freely on $H$.  Then, as a special case of the
decomposition above, we have that
\begin{equation}  \label{eq:wzw:decomp:1}
\left[ {\rm WZW}(H) / \Gamma \right]
\: = \:
\coprod_{\theta \in \hat{K}} {\rm WZW}( H/G )_{\theta(\omega)},
\end{equation}
with both sides at the same level.  That said, (ordinary) 
discrete torsion vanishes for cyclic subgroups, so the only occasion on which
$\theta(\omega)$ can be nontrivial will be if $H = {\rm Spin}(4n)$
and $\Gamma/K = {\mathbb Z}_2 \times {\mathbb Z}_2$.
(We will discuss that case in section~\ref{sect:ex:cs-spin-4n}.)

For example, consider a ${\mathbb Z}_4$ orbifold of an $SU(2)$ WZW model,
where a ${\mathbb Z}_2 \subset {\mathbb Z}_4$ acts trivially,
and the ${\mathbb Z}_2$ coset is the freely-acting center of $SU(2)$.
For an ordinary (orientation-preserving) orbifold, since
there is no discrete torsion in a ${\mathbb Z}_2$ orbifold, we have that
\begin{equation}
\left[ {\rm WZW}(SU(2)) / {\mathbb Z}_4 \right]
\: = \:
\coprod_{2} {\rm WZW}(SO(3))
\end{equation}
(with all WZW models at the same level).

Although we will not utilize orientifolds in this paper, in principle
one
can also consider orientation-reversing orbifolds (orientifolds) of 
WZW models, see 
e.g.~\cite{Gawedzki:2007uz,Brunner:2001fs,Pradisi:1995qy,Pradisi:1995pp,Schreiber:2005mi,Bachas:2001id}.
See \cite{Sharpe:2009hr} and references therein for discussions of
discrete torsion in orientifolds.

So far we have discussed discrete torsion weighting different
universes.  In principle, WZW models can also be weighted by
analogues of discrete theta angles.  Although these are better known
in the case of gauge theories\footnote{
Discrete theta angles in gauge theories in unrelated contexts have a long
history, see e.g.~\cite{Hori:1994uf}, \cite[section 6]{Hori:1994nc},
\cite[section 4]{Hori:2011pd} for two-dimensional examples and
 \cite{Gaiotto:2010be,Aharony:2013hda,Ang:2019txy}
for four-dimensional examples.
}, the point is that if
a group manifold $G$ has a torsion characteristic class,
some $w \in H^2(G,F)$ for some coefficient module $F$,
then there exists a discrete theta angle $\theta \in \hat{F}$
that weights maps into $G$, via a term in the action of the form
\begin{equation}
\int_{\Sigma} \langle \theta, \phi^* w \rangle,
\end{equation}
where $\Sigma$ is the worldsheet and $\phi: \Sigma \rightarrow G$
any map in the path integral.  If $G = \tilde{G}/Z$ for some finite
group $Z$, these discrete theta angles can also, for appropriate 
$w$, correspond to choices of discrete torsion in a $Z$ orbifold of a
WZW model on $\tilde{G}$.

In section~\ref{sect:boundary-wzw} we shall see that the choices
of discrete theta angle above that arise in the WZW orbifolds appearing
on boundaries of decompositions of one-form-gauged Chern-Simons theories,
are the same as choices of discrete torsion.

\section{Decomposition in noneffective one-form symmetry gaugings}
\label{sect:decomp}

In general terms, one expects a decomposition in a $d$-dimensional
quantum field theory whenever it has a global $(d-1)$-form symmetry
\cite{Tanizaki:2019rbk,Cherman:2020cvw}.

A typical example of a decomposition in two dimensions involves
gauging a non-effective group action:  a group action in which a subgroup
acts trivially on the theory being gauged, in the sense that its generator
commutes with the operators of that theory:  $[J,{\cal O}] = 0$.
Gauging a trivially-acting group results in a global one-form symmetry,
which is responsible for a decomposition.

In principle, an analogous phenomeon exists in three dimensions,
involving the gauging of `trivially-acting' one-form symmetries.  
Here, for a one-form action
to be trivial means that it commutes with the line operators in the theory,
as we shall elaborate below.

In this section, after a short overview of the notion of
non-effective one-form symmetries, we make a precise prediction for
decomposition.

\subsection{Non-effective one-form symmetry group actions}
\label{sect:trivacting}

We define a `trivially-acting' one-form symmetry in terms of the
fusion algebra of the corresponding lines, and a 
`non-effective' one-form symmetry is one in which a subset of the
lines acts trivially.

First, let us recall some basics of gauging one-form symmetries,
which in three dimensions we will describe by the fusion algebra
of line operators (see e.g.~\cite[section 3.1]{Roumpedakis:2022aik},
\cite{js,sp}, \cite[section II]{Barkeshli:2014cna} 
and references therein for a detailed discussion), with gauging as in
e.g.~\cite{Moore:1989yh}.
Anomalies in such a gauging are discussed in
e.g.~\cite[section 2.3]{Benini:2022hzx}, \cite[section 2.1]{Hsin:2018vcg},
\cite{Moore:1988qv,bk,Kitaev:2005hzj}.
In order to be gaugeable, its 't Hooft anomaly must vanish,
which requires that the lines be mutually transparent,
meaning that they have trivial mutual braiding.
In particular, a one-form symmetry necessarily has abelian lines, for which
the braiding is completely characterized by their spins (see
e.g.~\cite[equ'n (2.28)]{Benini:2022hzx}, \cite[section 2]{Hsin:2018vcg}),
schematically
\begin{equation}
\raisebox{-25pt}{
\begin{picture}(30,50)
\ArrowLine(15,0)(15,50)  \Text(18,45)[l]{$b$}
\ArrowArc(15,25)(10,110,270)
\ArrowArc(15,25)(10,-90,70)
\Text(30,25)[l]{$a$}
\end{picture}
}
\: \: = \:
B(a,b)
\:
\raisebox{-25pt}{
\begin{picture}(10,50)
\ArrowLine(5,0)(5,50)
\Text(8,25)[l]{$b$}
\end{picture} 
}
\end{equation}
where
\begin{equation}
B(a,b) \: = \:
\exp\left( 2\pi i\left( h(a \times b) - h(a) - h(b) \right) \right),
\end{equation}
where $a$, $b$ denote lines, and
$h(a) \mod 1$ is the spin of the line $a$.
Note that if the spins are integers, then $B=1$ and there is no obstruction.
Conversely, if $B = 1$, then spins are integers or half integers.

We take\footnote{
We are using ``$B$'' to mean several different things in this section.
We use $BK$ to denote a one-form symmetry, a standard notation 
in mathematics, going back decades.  (In physics, the notation
$K^{[1]}$ is sometimes used instead.)  Later we will use
$BG$ to denote a classifying space.  In this section, we also
use $B(a,b)$ to denote line monodromies.
} a `trivially-acting $BK$' to be described by
a set of lines $\{ g \}$ such that all
other lines $b$ both
\begin{enumerate}
\item have trivial monodromy under $g$, meaning $B(g,b) = 1$, and also are
\item invariant under fusion
with $g$, $g \times b = b$,
\end{enumerate}
for all $g$.
(In effect, there are two conditions in three dimensions,
whereas invariance in two dimensions really boils down to a single constraint
of the form $[J, {\cal O}] = 0$.)

To be clear, this notion can be somewhat counterintuitive.
Consider for example $SU(2)$ Chern-Simons theory.
This theory has a $B {\mathbb Z}_2$ one-form symmetry defined by
the center of $SU(2)$.  However, although the classical action is invariant
under the center, the Wilson lines are not invariant,
as the $B {\mathbb Z}_2$ action multiplies Wilson lines by phases
(corresponding to the $n$-ality of the corresponding representation with
respect to the center).  In particular, the $B {\mathbb Z}_2$ action
on $SU(2)$ Chern-Simons theory defined by the center of $SU(2)$ is
not trivial.

\subsection{Basic decomposition prediction}

In \cite{Pantev:2022kpl}, it was argued that in a quotient by a 
2-group $\Gamma$ of the form
\begin{equation}
1 \: \longrightarrow \: BK \: \longrightarrow \: \Gamma \:
\longrightarrow \: G \: \longrightarrow \: 1,
\end{equation}
where the $BK$ acts trivially, the path integral sums over
both $K$ gerbes and a subset of $G$ bundles, specifically $G$ bundles
satisfying a constraint.

In general, if one has a group $H$ and an abelian group $A$ with a map
$d: A \rightarrow H$ whose image is in the center of $H$,
then the crossed module\footnote{
See appendix~\ref{app:crossed-module} for an introduction to
crossed modules, or alternatively
\cite[appendix A]{Lee:2021crt}, \cite[section 2]{Bhardwaj:2021wif}.
} $\Gamma_{\cdot} = \{A \rightarrow H\}$ 
defines a 2-group we shall label $\Gamma$.
So long as we are interested in flat bundles, we can apply the same
analysis as \cite{Pantev:2022kpl}, 
and argue that $\Gamma$ bundles on a three-manifold
$M$ map to $G = H/{\rm im}\, A$ bundles satisfying a condition.
This 2-group fits into an exact sequence
\begin{equation}  \label{eq:a-to-h}
1 \: \longrightarrow \: K \: \longrightarrow \: A \: 
\stackrel{d}{\longrightarrow} \: H \: \longrightarrow \: G
\: \longrightarrow \: 1,
\end{equation}
where $K = {\rm Ker}\, d$.  (Physically, $d$ just encodes the $A$
action, by projecting it to a subgroup of the center of $H$.)
This exact sequence defines an element
\begin{equation}
\omega \: \in \: H^3_{\rm group}(G,K) \: = \: H^3_{\rm sing}(BG,K),
\end{equation}
which we will give explicitly in~(\ref{eq:defn-omega}), 
and the condition that $G$ bundles must satisfy
to be im the image of $\Gamma$ bundles is that 
\begin{equation}  \label{eq:constr}
\phi^* \omega \: = \: 0,
\end{equation}
for $\phi: M \rightarrow B \Gamma$ the map defining the $\Gamma$ bundle
on $M$, for the same reasons discussed in \cite{Pantev:2022kpl}.

Next, we describe the element $\omega$ corresponding to the
extension~(\ref{eq:a-to-h}), appearing in the constraint~(\ref{eq:constr})
above.
Let $Z = {\rm im}\, d \subset Z(H)$, the center of $H$,
and $w_G$ the $Z$-valued degree-two characteristic class for
$G$ corresponding to a generator of
$H^2_{\rm sing}(BG,Z)$.  (For example, for $G = SO(n)$,
$w_G$ is the second Stiefel-Whitney class $w_2$.)  
Let $\alpha \in H^2_{\rm group}(Z,K)$
be the class of the extension
\begin{equation} \label{eq:ext-alpha}
1 \: \longrightarrow K \: \longrightarrow \: A
\: \longrightarrow \: Z \: \longrightarrow \: 1,
\end{equation}
and let
\begin{equation}
\beta_{\alpha}: \: H^2_{\rm sing}(BG,Z) \: \longrightarrow \:
H^3_{\rm sing}(BG,K)
\end{equation}
be the Bockstein homomorphism in the long exact sequence
associated to the extension~(\ref{eq:ext-alpha}).
Then
\begin{equation}  \label{eq:defn-omega}
\omega \: = \: \beta_{\alpha}( w_G ) \: \in \: H^3_{\rm sing}(BG,K).
\end{equation}

When discussing boundary WZW models, it will be useful to describe
$\omega$ differently.  To that end, we use the fact
that
\begin{equation}
H^n_{\rm sing}(BG,Z) \: = \: {\rm Map}\left( BG, K(Z,n) \right),
\end{equation}
to write write $w_G$ and $\alpha$ as maps
\begin{equation}
w_G: \: BG \: \longrightarrow \: K(Z,2), 
\: \: \:
\alpha: \: BZ ( = K(Z,1)) \: \longrightarrow \: K(K,2).
\end{equation}
Since Eilenberg-Maclane spaces are in the stable category,
where $B$ exists as a functor, we can define
\begin{equation}
B \alpha: \: K(Z,2) \: \longrightarrow \: K(K,3),
\end{equation}
hence
\begin{equation}
B\alpha \circ w_G: \: BG \: \longrightarrow \: K(K,3),
\end{equation}
and so defines an element of $H^3_{\rm sing}(BG,K)$.
Furthermore, $B \alpha$ is just the Bockstein homomorphism $\beta_{\alpha}$,
hence
\begin{equation}  \label{eq:balpha-bock}
\omega \: = \:
B \alpha \circ w_G \: = \: \beta_{\alpha}(w_G),
\end{equation}
and so we recover the description of $\omega$ above.

So far, we have argued that on general principles,
our $\Gamma$ gauge theory should be described by a $G$ gauge theory
such that the $G$ bundles satisfy the constraint~(\ref{eq:constr}).
Just as in \cite{Hellerman:2006zs,Pantev:2022kpl}, 
such a restriction on instantons
can be implemented by a sum over universes.  
The constraint~(\ref{eq:constr}), namely $\phi^* \omega = 0$,
is implemented by summing over $G$ Chern-Simons theories
with discrete theta angles coupling to $\omega$, formally
\begin{equation}  \label{eq:decomp-predict}
\left[ \mbox{Chern-Simons}(H) / B A \right]
 \: = \: \coprod_{\theta \in \hat{K} }
\mbox{Chern-Simons}(G)_{\theta},
\end{equation}
where $\theta$ is the three-dimensional discrete theta angle
coupling to $\phi^* \omega$, for levels and underlying three-manifolds
for which these theories are defined\footnote{
As has been noted in e.g.~\cite{Moore:1989yh}, 
\cite[appendix C]{Seiberg:2016rsg},
\cite[appendix A]{Seiberg:2016gmd}, 
\cite{Belov:2005ze,Freed:1992vw,freed2},
not every Chern-Simons theory with every
level is well-defined on every three-manifold.
The basic issue is that Chern-Simons actions are not precisely 
gauge-invariant, but under gauge transformations shift by an amount
proportional to $2\pi$.  Depending upon the gauge group and the three-manifold,
the proportionality factor may or may not be integral.  If $k$ times that
proportionality factor is integral, then the exponential of the action
is gauge-invariant, and the theory is well-defined; if that product is not
integral, then the path integral is not gauge-invariant and so not defined.
Even if it is defined, it may depend upon subtle choices.
For example, \cite[appendix A]{Seiberg:2016gmd} argues that the
(ordinary, bosonic)
$U(1)_1$ Chern-Simons theory is well-defined only on spin three-manifolds,
and furthermore that the choices of values of the action, the
Chern-Simons invariants in the sense of \cite{Borel:1999bx,deBoer:2001wca}, 
are in one-to-one
correspondence with the spin structures.
More generally, gauging one-form symmetries can create issues of this form,
precisely because one twists gauge fields by gerbes, which results in 
`twisted' bundles and connections not present in the original theory, of
fractional instanton numbers.
}.
This is our prediction for decomposition in three-dimensional Chern-Simons
theories.

The $G$ Chern-Simons theory
is defined to be the $B({\rm im}\,A)$ gauging of the $H$ Chern-Simons theory,
at the same level as the $H$ Chern-Simons theory.  This is important
to distinguish because
sometimes gauging one-form symmetries can shift levels.  For example,
\cite[section C.1]{Seiberg:2016rsg} argues that, schematically,
$U(1)_{4m}/B{\mathbb Z}_2 = U(1)_m$, and not $U(1)_{4m}$,
despite the fact that as groups, $U(1) / {\mathbb Z}_2 = U(1)$.

The reader should note that the decomposition statement above
correctly reproduces
ordinary one-form gaugings. 
Consider the case that $K = 1$, so that the map $d: A \rightarrow H$
is one-to-one into the center of $H$.  
Then, decomposition~(\ref{eq:decomp-predict}) correctly predicts that
\begin{equation}
\left[ \mbox{Chern-Simons}(H) / BA \right] \: = \: \mbox{Chern-Simons}(G),
\end{equation}
which is a standard result (see e.g.~\cite{Moore:1989yh}).  
Decomposition becomes interesting
in cases in which $K \neq 1$.

In section~\ref{sect:exs} we will check this statement in several
examples, outlining how it both reproduces known results as well as
explains new cases.

\subsection{Boundary WZW models}
\label{sect:boundary-wzw}

Let us now turn to Chern-Simons theories on manifolds with
boundary, and the corresponding theories on the boundaries.
We will see that the bulk Chern-Simons decomposition of the previous
section correctly predicts
a decomposition of boundary WZW models, which matches existing
results on decomposition in two-dimensional orbifolds.
This matching involves a rather interesting relation
between characteristic classes of bundles on three-manifolds
and choices of discrete torsion in two-dimensional orbifolds.
In particular, the fact that the three-dimensional decomposition
correctly reproduces two-dimensional decomposition on
the boundary
is an important consistency test of our proposal.

Briefly, as has been discussed elsewhere 
(see 
e.g.~\cite{Elitzur:1989nr,Bos:1989kn,Axelrod:1989xt,Coussaert:1995zp,ncat}, 
\cite[section 4.2]{Dijkgraaf:1989pz}, 
\cite[section 5.2]{Gawedzki:1999bq},
and in related contexts
\cite{Fiorenza:2012ec,frs}),
on a three-manifold with boundary,
a bulk Chern-Simons theory for gauge group $G$
naturally couples to a (chiral) WZW model for the group $G$
on the boundary.  If the Chern-Simons theory has level $k$,
then (see e.g.~\cite[section 4.2]{Dijkgraaf:1989pz}) the boundary
WZW model has level $\tau(k)$, where
\begin{equation}
\tau: \: H^n_{\rm sing}(BG,F) \: \longrightarrow \: H^{n-1}_{\rm sing}(G,F)
\end{equation}
is the loop space map\footnote{
This is the natural map
\begin{eqnarray}
\lefteqn{
H^n_{\rm sing}(BG,F) \: = \: {\rm Map}\left( BG, K(F,n) \right)
} \nonumber \\
& \hspace*{0.25in} \longrightarrow &
{\rm Map}\left( \Omega( BG ), \Omega( K(F,n) ) \right) \: = \:
{\rm Map}\left( G, K(F,n-1) \right) \: = \: H^{n-1}_{\rm sing}(G,F).
\end{eqnarray} 
which sends any $f \in {\rm Map}(BG, K(F,n))$ to $\Omega(f)$.
For later use, to construct explicit maps, one needs concrete choices of
e.g.~$X \mapsto \Omega B X$, for which we refer the reader to
e.g.~\cite{segal1,dunn1,may1,maythomas}.  As such choices do not alter
cohomology classes, we will not discuss them explicitly in this paper.
}
for any abelian group $F$, and we take Chern-Simons levels\footnote{
As before, levels are assumed to be such that the theory exists.
} 
$k \in H^4_{\rm sing}(BG,{\mathbb Z})$, and WZW levels 
$\tau(k) \in H^3_{\rm sing}(G,{\mathbb Z})$. 
Similarly, if the Chern-Simons theory has a discrete theta angle
coupling to some characteristic class defined by an element of
$\omega \in H^3(BG,F)$, then the boundary WZW model couples\footnote{
We would like to thank Y.~Tachikawa for a discussion of discrete theta
angles in this context.  
} to a 
discrete theta angle defined by
$\tau(\omega) \in H^2(G,F)$.
Such discrete theta angles in two-dimensional WZW models
are reviewed in section~\ref{sect:review-wzw}.

Given that standard bulk Chern-Simons / boundary WZW model relationship
reviewed above,
the three-dimensional decomposition prediction~\ref{eq:decomp-predict}
implies that in the associated boundary RCFT,
an $A$ orbifold of a WZW model for $H$ is equivalent to a disjoint
union of WZW models for $G$,
\begin{equation}  \label{eq:wzw:decomp:boundary}
[ {\rm WZW}(H) / A ] \: = \: \coprod_{\theta \in \hat{K}}
 {\rm WZW}(G)_{\theta},
\end{equation}
with levels and discrete theta angles related to those of the
bulk theory by the map $\tau$.
We will see later in this section that although the WZW discrete
theta angles $\theta$
are derived from characteristic classes in the Chern-Simons
theory, they nevertheless correspond to choices of discrete torsion in the
boundary orbifolds.

As a consistency check, let us show that $\tau$ commutes with
gauging $BA$, so that the levels on the left and right-hand sides 
of~(\ref{eq:wzw:decomp:boundary}) match, just as they did\footnote{
Modulo subtleties discussed there in special cases,
such as those arising from the fact that
$U(1)/{\mathbb Z}_k = U(1)$ as a group, but the corresponding
Chern-Simons theories have different levels.
} in
the bulk prediction~(\ref{eq:decomp-predict}).
First, for $G$ any topological group, there is a natural homotopy
equivalence between the loop space $\Omega(BG)$ and $G$ (meaning
that $BG$ is a delooping of $G$).  Also, for any abelian group $F$,
the Eilenberg-Maclane space $K(F,n-1)$ is homotopy equivalent to 
loop space $\Omega( K(F,n) )$.  Since
\begin{equation}
H^n_{\rm sing}(BG,F) \: = \: {\rm Map}(BG, K(F,n))
\end{equation}
and since $\Omega$ is a functor, 
for any continuous homomorphism $f: G_1 \rightarrow G_2$ between
topological groups $G_1$, $G_2$, there is a continuous map
$Bf: B G_1 \rightarrow B G_2$ and natural maps
\begin{eqnarray}
{\rm Map}\left( BG_2, K(F,n) \right) & \longrightarrow & 
{\rm Map}\left( BG_1, K(F,n) \right),
\\
a & \mapsto & B(f \circ a)
\end{eqnarray}
and
\begin{eqnarray}
{\rm Map}\left(G_2, K(F,n-1) \right) & \longrightarrow &
{\rm Map}\left(G_1, K(F,n-1) \right),
\\
b & \mapsto & f \circ b.
\end{eqnarray}
Combining these maps, one finds that for any Lie group $G$ with
$K$ a subgroup of the center, the following diagram commutes:
\begin{equation}
\xymatrix{
H^3_{\rm sing}(B (G/K), F) \ar[r] \ar[d] & H^2_{\rm sing}( G/K, F) \ar[d]
\\
H^3_{\rm sing}(BG, F) \ar[r] & H^2_{\rm sing}(G,F).
}
\end{equation}
This tells us that the levels appearing on either side of
the boundary WZW relation~(\ref{eq:wzw:decomp:boundary}) match,
as expected, consistent with the prediction~(\ref{eq:decomp-predict})
of the bulk Chern-Simons theory.

Now, we will argue that the WZW model discrete theta angles,
arising as $\tau$ of characteristic classes in the Chern-Simons theory,
are the same as choices of discrete torsion in the boundary theory. 
This will be important in understanding how the three-dimensional
Chern-Simons decomposition compares to two-dimensional decompositions
as reviewed in section~\ref{sect:review-wzw}.
For simplicity, we will assume that $H$ is the universal covering,
so that $Z = \pi_1(G)$.  (Similar results exist in more general cases.)

To that end, since $\tau$ is the loop space functor, we can write
\begin{equation}
\tau\left( \beta_{\alpha}(w_G) \right) \: = \:
\tau( B \alpha \circ w_G ) \: = \: \Omega(B \alpha \circ w_G)
\: = \: \Omega(B \alpha) \circ \Omega(w_G).
\end{equation}
Now, $\Omega(B \alpha) = \alpha$, and
\begin{equation}
\Omega(w_G) \: \in \: {\rm Map}( \Omega(BG), \Omega( K(Z,2) ) )
\: = \: {\rm Map}(G, K(Z,1) ),
\end{equation}
so $\Omega(w_G)$ is a map $G \rightarrow BZ$.
Now, we claim that $\Omega(w_G)$ is also the cell attachment map $p$
of the Postnikov tower,
$p: G \rightarrow B \pi_1(G) = BZ$, where $Z = \pi_1(G)$.

To make this clear, recall that the Postnikov tower map
is the classifying map for the universal cover.  In other words,
if $\tilde{G}$ is the universal covering group of $G$, then
$p^* EZ = \tilde{G}$.  Now, on the other hand,
$B \tilde{G} \rightarrow BG$ is a principal $K(Z,1)$ bundle on $BG$,
which corresponds to a map $BG \rightarrow B( K(Z,1) ) = K(Z,2)$,
which is $w_G$.  Applying the loop space functor gives
the map $\Omega(w_G): G \rightarrow K(Z,1) = BZ$, which is then more or
less tautologically $p$.

In particular, we see that
\begin{equation}
\tau( B \alpha \circ w_G) \: = \: 
\alpha \circ \Omega(w_G) \: = \: \alpha \circ p.
\end{equation}

The expression above relates the Chern-Simons discrete
theta angles (coupling to bundle characteristic classes)
to discrete torsion on the boundary.
We can see this as follows.
If $\phi: \Sigma \rightarrow G$ is any map from the
worldsheet $\Sigma$ into the target $G$, then
$p \circ \phi: \Sigma \rightarrow BZ$ defines a $Z$-twisted sector
over $\Sigma$.  In particular, the discrete theta angle phase
\begin{equation}
\langle \theta, \phi^* (\alpha \circ p) \rangle,
\end{equation}
for for $\theta: K \rightarrow U(1)$
any character of $K$,
corresponds to discrete torsion in the $Z$-twisted sector defined by
$p \circ \phi$, specifically discrete torsion given by
$\theta(\alpha) \in H^2_{\rm group}(Z, U(1))$, for
$\alpha \in H^2_{\rm group}(Z,K)$.
Thus, we see that $\tau$ relates discrete theta angles coupling
to bundle characteristic classes on three-manifolds,
to discrete torsion in two-dimensional orbifolds on boundaries.

In passing, this phenomenon that three-dimensional bulk discrete
theta angles become discrete torsion in boundary two-dimensional
orbifolds is also visible in the case that the bulk theory is a finite
2-group orbifold, see \cite[section 3.2]{Pantev:2022kpl}.

Now, let us compare the decomposition~(\ref{eq:wzw:decomp:boundary}) 
in boundary WZW models,
implied by bulk Chern-Simons decomposition, to 
standard results \cite{Hellerman:2006zs} on decomposition in two-dimensional
orbifolds, as reviewed earlier in section~\ref{sect:review-wzw}.

Certainly the form of the boundary decomposition~(\ref{eq:wzw:decomp:boundary})
is identical to that arising in two-dimensional orbifolds with
trivially-acting central subgroups, possibly modulo the form of the discrete
theta angles.  We have just argued that the discrete theta angles
arising on the boundary correspond to choices of discrete torsion,
and in fact, the discrete torsion phases arising in the boundary
case match those in the ordinary two-dimensional case.

We can relate these two pictures of boundary discrete theta angles as follows.
Recall $\alpha \in H^2_{\rm group}(Z,K)$ is the class of the extension
\begin{equation}
1 \: \longrightarrow \: K \: \longrightarrow \: A \: \longrightarrow \:
Z \: \longrightarrow \: 1.
\end{equation}
In two-dimensional decomposition in $A$ orbifolds with trivially-acting
central subgroups $K$, the discrete torsion phase factors on a
universe associated with $\theta \in \hat{K}$ are precisely the
image of $\alpha$ under $\theta$:
\begin{eqnarray}
H^2_{\rm group}(Z,K) & \longrightarrow & H^2_{\rm group}(Z,U(1)),
\\
\alpha & \mapsto & \theta \circ \alpha.
\end{eqnarray}
These are the same as the discrete torsion phases arising in the boundary
WZW decomposition~(\ref{eq:wzw:decomp:boundary}), as we have just
discussed, and we will confirm explicitly in
examples in section~\ref{sect:exs}
that the decomposition above
in the boundary theory precisely coincides with the decomposition
of WZW orbifolds given in~(\ref{eq:wzw:decomp:1}).
This matching is an important consistency test of our proposal.

\subsection{Nontriviality of discrete theta angles}
\label{sect:nontriv}

In the boundary WZW models appearing in these decompositions,
the
discrete torsion on each universe appearing in a decomposition
is trivial.  For most single group factors, this is because the
center is usually a cyclic group, and cyclic group orbifolds have
no discrete torsion.  The exceptions are the groups Spin$(4n)$,
which have center ${\mathbb Z}_2 \times {\mathbb Z}_2$.
That finite group admits discrete torsion; however, to generate
the discrete torsion in a decomposition of a string orbifold,
the orbifold group must be nonabelian, and so cannot arise as the
boundary of a three-dimensional theory, as we will discuss
in greater detail in section~\ref{sect:ex:cs-spin-4n}.

In at least some examples, not only are the boundary discrete theta
angles (discrete torsions) trivial, but the bulk discrete theta angles
are also trivial.
For example\footnote{
We would like to thank Y.~Tachikawa for making this observation.
}, 
in bulk theories, for cases in which $K = {\mathbb Z}_2$, $Z = {\mathbb Z}_2$,
and $A = {\mathbb Z}_4$, so that the extension $\alpha$ is
\begin{equation}
1 \: \longrightarrow \: {\mathbb Z}_2 \: \longrightarrow {\mathbb Z}_4
\: \longrightarrow \: {\mathbb Z}_2 \: \longrightarrow \: 1,
\end{equation}
the bulk discrete theta angle couples to the Bockstein $\beta_{\alpha}$
of a distinguished
element $w_G \in H^2(M_3,{\mathbb Z}_2)$.  Now, for this $\alpha$,
\begin{equation}
\beta_{\alpha}(w_G) \: = \: {\rm Sq}^1(w_G),
\end{equation}
and as we will argue in section~\ref{sect:ex:su2-bz4},
\begin{equation}
{\rm Sq}^1(w_G) \: = \:  
w_1(TM_3) \cup w_G,
\end{equation} 
hence it can only be nonzero on nonorientable spaces.
However, we only define Chern-Simons theories on oriented three-manifolds,
so for all cases we consider, these bulk discrete theta angles vanish.

Similarly, if the three-manifold is $T^3$, the pertinent
Bockstein homomorphism will vanish, and one cannot get a nonzero bulk
discrete theta angle.  Briefly, for any short exact sequence
\begin{equation}
1 \: \longrightarrow \: K \: \longrightarrow \: A \: \longrightarrow \:
Z \: \longrightarrow \: 1,
\end{equation}
for $K, A, Z$ abelian, the induced map
\begin{equation}
H^2( T^3, A ) \: \longrightarrow \: H^2(T^3, Z)
\end{equation}
is surjective (since each of those cohomology groups is just Hom from
a free abelian group into the coefficients), 
which implies that in the long exact sequence
\begin{equation}
H^2(T^3, K) \: \longrightarrow \: H^2(T^3, A) \: 
\longrightarrow \: H^2(T^3, Z) \: 
\stackrel{\beta}{\longrightarrow} \: H^3(T^3, K),
\end{equation}
the Bockstein $\beta = 0$, and so the bulk discrete theta angles
are trivial in corresponding cases.

For another example, consider Lens spaces.
From \cite[example 3E.2]{hatcher}, for the Bockstein associated to the
short exact sequence
\begin{equation}
1 \: \longrightarrow \: {\mathbb Z}_m \: \longrightarrow \:
{\mathbb Z}_{m^2} \: \longrightarrow \: 
{\mathbb Z}_m \: \longrightarrow \: 1,
\end{equation}
the associated Bockstein maps generators of $H^1(L, {\mathbb Z}_m)$ to
generators of $H^2(L,{\mathbb Z}_m)$, for $L$ a Lens space,
but $\beta^2 = 0$, hence
the associated Bockstein map 
\begin{equation}
\beta: \: H^2(L, {\mathbb Z}_m) \: \longrightarrow \:
H^3(L, {\mathbb Z}_m)
\end{equation}
necessarily vanishes, and so the bulk discrete theta angles are trivial
in corresponding cases.

More generally, whether the bulk discrete theta angles are
always trivial is a reflection of the map $\tau: H^3_{\rm sing}(BG,K)
\rightarrow H^2_{\rm sing}(BG,K)$.  For example, 
if $\tau$ is injective, then triviality of the boundary discrete theta angles
implies triviality of the bulk discrete theta angles.  We leave 
general questions about the injectivity of $\tau$ for future work.
 
In passing, note 
that in the bulk, 
orientability plays a key role.
At least abstractly, it is tempting to speculate about
more general cases involving e.g.~orientifolds of
boundary WZW modelsi, as might arise if the three-manifold descends to
a solid Klein bottle (a three-manifold whose boundary is the two-dimensional
Klein bottle).  On such a nonorientable space, at least sometimes
the discrete theta angles would be nontrivial.  Furthermore,
in orientifolds, discrete torsion is counted by $H^2_{\rm group} (Z,U(1))$ with
a nontrivial action on the coefficients
(see e.g.~\cite{Sharpe:2009hr,Bachas:2001id,Gawedzki:2007uz,Brunner:2001fs}),
so that for example
$H^2_{\rm group}({\mathbb Z}_2,U(1))$ can be nonzero, which again would
result in boundary WZW models with nonzero discrete theta angle
contributions.

\section{Spectra}
\label{sect:spectra}

In this section we briefly describe the spectra of monopole operators
and line operators in a theory with a gauged trivially-acting one-form
symmetry, and argue that the results are consistent with
decomposition~(\ref{eq:decomp-predict}).

\subsection{Monopole operators}

In two dimensional theories, when one gauges a non-effectively-acting group,
one gets twist fields
and Gukov-Witten operators corresponding to conjugacy classes in the
trivially-acting subgroup.
In three dimensional theories, instead of twist fields,
one has monopole operators (see e.g.~\cite{Borokhov:2002ib,Borokhov:2002cg}),
which play the same role.
In this section we will outline their properties.

In two dimensions, twist fields generate branch cuts, which in the
language of topological defect lines are real codimension-one walls
that implement the gauging of the zero-form symmetry.
In three dimensions, when gauging a one-form symmetry, from thinking about
topological defect ones one sees the theory has codimension-two lines,
which end in monopole operators, in the same way that in two dimensions,
the orbifold branch cuts terminate in twist fields.

We can think of the monopole operators in three dimensions
as local disorder operators:
on a sphere surrounding the monopole operator associated to a $BG$ symmetry,
one has a nontrivial $G$ gerbe, corresponding to an element of $H^2(S^2,G)$
(for $G$ assumed finite), just as on a circle surrounding a twist field
in two dimensions one has a nontrivial bundle.

In two dimensions, the twist fields associated to trivially-acting gauged
zero-form symmetries are local dimension-zero operators, which can be
used to form projectors onto the universes of decomposition.
In three dimensions, the monopole operators associated to trivially-acting
gauged one-form symmetries are closely analogous, and can again be used to
form projection operators, in exactly the same fashion.
In \cite[section 4.1.4]{Pantev:2022kpl}, projection operators are explicitly
constructed from monopole operators in three-dimensional theories, 
and we encourage the reader to
consult that reference for further details.

\subsection{Line operator spectrum}
\label{sect:lineops}

Given a `gaugable' one-form symmetry, described by a subset of
the lines in the theory,
there is a standard procedure for computing the
spectrum of lines in the gauged theory, given as follows
(see e.g.~\cite[section 2]{Moore:1989yh},
\cite[section 2.5]{Cordova:2017vab}, \cite{Benini:2022hzx}).
For $B {\mathbb Z}_n$,
let $g$ denote a line generating the others, and then:
\begin{itemize}
\item Exclude from the spectrum all lines $a$ which have monodromy\footnote{
In terms of the S-matrix, $B(a,b) = S_{ab}/S_{0b}$,
see e.g.~\cite[equ'n (40)]{Barkeshli:2014cna}.
}
$B(g,a) \neq 1$,
under a generator $g$,
where the $g$ action on a line $b$ is determined by the process
\begin{equation}
\raisebox{-25pt}{
\begin{picture}(30,50)
\ArrowLine(15,0)(15,50)  \Text(18,45)[l]{$b$}
\ArrowArc(15,25)(10,110,270)
\ArrowArc(15,25)(10,-90,70)
\Text(30,25)[l]{$g$}
\end{picture}
}
\: \: = \:
B(g,b)
\:
\raisebox{-25pt}{
\begin{picture}(10,50)
\ArrowLine(5,0)(5,50)
\Text(8,25)[l]{$b$}
\end{picture} 
}
\end{equation}

\item Identify any two lines $b$, $g \times b$ that differ by fusion with $g$,
and finally
\item Lines $b$ that are invariant under fusion (meaning
$a = g \times a$)
become $n$ lines in the spectrum of the gauged
theory.
\end{itemize}
This is closely analogous to two-dimensional orbifolds, in which
one omits non-invariant operators, and fixed points lead to twist fields.

This is a special case of a more general procedure, sometimes known
as anyon condensation, which is also applicable to noninvertible symmetries,
unlike the basic algorithm above.
See e.g.~\cite[section 4.1]{Benini:2022hzx},
\cite{Kong:2013aya,Fuchs:2002cm,Cordova:2017vab,Yu:2021zmu,Gaiotto:2019xmp} 
for further details.

Now, in our case, the (noneffectively-acting)
one-form symmetry we wish to gauge is not described
by a set of lines within the original theory.  Sometimes, in some special
cases, we can describe it by adding lines to the theory, such as in the 
case that the entire gauged one-form symmetry acts trivially.
In general, however, that procedure is not well-defined.
Consider for example the case of $SU(2)_4$ Chern-Simons theory,
whose spectrum of lines $\{ (0), (1), (2), (3), (4) \}$
is as described in
appendix~\ref{app:lineops}.  Let us consider gauging a
$B {\mathbb Z}_4$.  Now, $SU(2)_4$ has a $B {\mathbb Z}_2$ symmetry,
coresponding to the lines (0), (1), so one could imagine extending it
to $B {\mathbb Z}_4$
by replacing $\{ (0), (1) \}$ with
$\{ (0), \ell_1, \ell_2, \ell_3 \}$
which obey
\begin{equation}
\ell_i \times \ell_j \: = \: \ell_{i + j \mod 4},
\end{equation}
with $\ell_0 = (0)$.  In order for the image of ${\mathbb Z}_2 \hookrightarrow
{\mathbb Z}_4$ to act trivially, we require
\begin{equation}
\ell_2 \times (2,3,4) \: = \: (2,3,4),
\end{equation}
and for this to descend to the ordinary $SU(2)_4$, we also require
\begin{equation}
\ell_1 \times (2,3,4) \: = \: \ell_3 \times (2,3,4),
\end{equation}
which must match $(1) \times (2,3,4)$ in the $SU(2)_4$ fusion algebra
given in appendix~\ref{app:lineops}:
\begin{equation}
\ell_{1,3} \times (2) \: = \: (3), \: \: \:
\ell_{1,3} \times (3) \: = \: (2), \: \: \:
\ell_{1,3} \times (4) \: = \: (4).
\end{equation}
The new lines $\ell_{i}$ are then defined to have trivial monodromy with
all other lines
\begin{equation}
B(\ell_i, x) \: = \: 1
\end{equation}
for all lines $x$.
This much is uniquely specified by the statement of the extension.

Now, we have not completely specified the extension of $SU(2)_4$;
for example, the product $(2) \times (3)$ in $SU(2)_4$ contains a (1),
so we would still need to decide whether to replace (1) with $\ell_1$ or
$\ell_3$, for example.  However, in making such choices,
we find an internal contradiction with the structure we have already
described, namely a failure of associativity.
For example, in $SU(2)_4$, as described in appendix~\ref{app:lineops},
\begin{equation}
(2) \times (3) \: = \: (1) + (4), \: \: \: 
(3) \times (3) \: = \: (0) + (4).
\end{equation}
We could replace the (1) above with either $\ell_1$ or $\ell_3$.
Suppose we take 
\begin{equation}
(2) \times (3) \: = \: \ell_1 + (4).
\end{equation}
Then,
\begin{equation}
\ell_1 \times ( (2) \times (3) ) \: = \: \ell_1 \times (\ell_1 + (4) )
\: = \: \ell_2 + (4),
\end{equation}
\begin{equation}
( \ell_1 \times (2) ) \times (3) \: = \: (3) \times (3) \: = \: (0) + (4).
\end{equation}
However, $\ell_2 \neq (0)$, so we see that
\begin{equation}
\ell_1 \times ( (2) \times (3) )
\: \neq \:
( \ell_1 \times (2) ) \times (3),
\end{equation}
and so associativity is broken.  We encounter a similar problem if we
choose
\begin{equation}
(2) \times (3) \: = \: \ell_3 + (4)
\end{equation}
instead and consider fusion with $\ell_3$.  
Put simply, we cannot enlarge the $B {\mathbb Z}_2$ inside
$SU(2)_4$ to a noneffective $B {\mathbb Z}_4$, without breaking
associativity.

With this in mind, we outline here\footnote{
Although we are only interested in isomorphism classes of objects,
presumably the full categorical description is in terms of module categories,
or 
as a minor variation on the group actions
described in \cite[section III.B]{Barkeshli:2014cna}.  
As we are only interested here in counting 
(isomorphism classes of) objects, and this is merely a minor
variation on existing methods, we will be very schematic.
} a minor extension of the prescription of
\cite[section 2]{Moore:1989yh},
\cite[section 2.5]{Cordova:2017vab}, \cite{Benini:2022hzx} for counting
line operators in three-dimensional theories with gauged one-form symmetries.
(As we will not be using noninvertible symmetries, we will not attempt
to describe the analogous construction for condensation algebra objects
here.)

Our approach is motivated by the action of a group $G$ on a set $M$:
we distinguish $G$ and $M$, $G$ is not a subset of $M$ in general,
though we can still define an action of $G$ on $M$ that enables us to
make sense of the quotient $M/G$.
Let $G$ be a finite abelian group, so that $BG$ is a group of one-form
symmetries, and associate lines to elements of $G$.  Consider a set of
simple lines ${\cal C}$ (objects in a braided tensor category, which we will
gauge).  

An action of $BG$ on ${\cal C}$ is then described by giving, for each
$g \in G$ and line $L \in {\cal C}$,
\begin{itemize}
\item a monodromy $B(g,L)$, such that
\begin{equation}
B(g_1, L) \, B(g_2, L) \: = \: B(g_1 g_2, L),
\end{equation}
and
\item a fusion $g \times L \in {\mathbb Z}[{\cal C}]$,
meaning $g \times L = \sum_c N^c_{gL} c$ for $c \in {\cal C}$ and
$N^c_{gL} \in {\mathbb Z}$, with the property that
\begin{equation}
g_1 \times \left( g_2 \times L \right) \: = \: (g_1 g_2) \times L.
\end{equation}
\end{itemize}
We say that a line in $BG$ corresponding to $g \in G$ acts trivially
if,
for all $L \in {\cal C}$, 
\begin{equation}
B(g,L) \: = \: 1 \: \: \: \mbox{ and } \: \: \:
g \times L \: = \: L,
\end{equation}
and then to say $BG$ acts noneffectively, as in section~\ref{sect:trivacting},
means that some for some $g \neq 1$ in $G$, the line corresponding to
$G$ acts trivially.

Now, given an action of $BG$ on ${\cal C}$,
we propose to construct the lines of the quotient ${\cal C}/BG$ as follows,
by close analogy with \cite[section 2]{Moore:1989yh},
\cite[section 2.5]{Cordova:2017vab}, \cite{Benini:2022hzx}.
\begin{itemize}
\item Exclude any $L$ such that for some $g \in G$, $B(g,L) \neq 1$,
\item Identify $L \sim g \times L$ for each $g \in G$,
\item For each $g \in G$ such that $g \times L = L$, we get a copy of $L$
in ${\cal C}/BG$.
\end{itemize}
It is straightforward to check that in the special case 
the lines in $B G$ are a subset of those in ${\cal C}$, this reduces
to the prescription reviewed earlier and in 
\cite[section 2]{Moore:1989yh},
\cite[section 2.5]{Cordova:2017vab}, \cite{Benini:2022hzx}.

As another special case, note that if all of $BG$ acts trivially, then
in the quotient ${\cal C}/BG$,
\begin{itemize}
\item No lines in ${\cal C}$ are excluded, since $B(g,L) = 1$ for all
$L$,
\item No lines are identified, since $g \times L = L$ for all $L$,
so fusion does not relate different lines,
\item Since $g \times L = L$ for each $g \in G$ and each $L \in
{\cal C}$, we get $|G|$ copies of the lines in ${\cal C}$.
\end{itemize}
This is consistent with the expectations of decomposition in this case:
if we gauge a $BG$ that acts completely trivially on a theory,
in the sense above, one expects to get $|G|$ copies of the theory.

We will apply this computation in specific examples in Chern-Simons
theories later in this paper, but for the moment, we give two
toy examples, to illustrate the idea.

First, consider $B {\mathbb Z}_2 / B {\mathbb Z}_2$.
Let the lines of ${\cal C} = B {\mathbb Z}_2$ be generated over ${\mathbb Z}$
by $\{ (0), (1) \}$, where
\begin{equation}
(0) \times (0) \: = \: (0), \: \: \:
(0) \times (1) \: = \: (1), \: \: \:
(1) \times (1) \: = \: (0),
\end{equation}
and $B {\mathbb Z}_2$ acts as
\begin{equation}
g \times (0) \: = \: (1), \: \: \:
g \times (1) \: = \: (0),
\end{equation}
and we take all monodromies $B = 1$.
Then, applying the procedure above, to get the lines of ${\cal C}/B {\mathbb Z}_2$,
\begin{itemize}
\item Since $B(g,L) = 1$ for all $L \in {\cal C}$, no lines are excluded,
\item Since $g \times (0) = (1)$, $(0) \sim (1)$,
\item No lines are invariant.
\end{itemize}
Hence the quotient is generated by one single line, as one would expect.

Next, consider  $B {\mathbb Z}_2 / B {\mathbb Z}_4$, where
the $B {\mathbb Z}_4$ acts noneffectively.
Let the lines of ${\cal C} = B {\mathbb Z}_2$ be as above,
and the generator $g$ of $B {\mathbb Z}_4$ acts as
\begin{equation}
g \times (0) \: = \: (1), \: \: \:
g \times (1) \: = \: (0).
\end{equation}
As before, we take all monodromies $B = 1$.
Applying the procedure above,
\begin{itemize}
\item Since $B(g,L) = 1$ for all $L \in {\cal C}$, no lines are excluded.
\item Since $g \times (0) = (1)$, $(0) \sim (1)$,
\item Since $g^2 \times (0) = (0)$ and $g^2 \times (1) = (1)$,
$(0) \sim (1)$ appears twice in the quotient.
\end{itemize}
Thus, the quotient ${\cal C} / B {\mathbb Z}_4$ is generated over
${\mathbb Z}$ by two lines, as expected since a ${\mathbb Z}_2 \subset
{\mathbb Z}_4$ describes trivially-acting lines.

We should also briefly observe that the theories we are describing,
which decompose, have the property that they violate the
axiom of remote detectability in a topological order,
see e.g.~\cite{Levin:2013gaa,Kong:2014qka,Johnson-Freyd:2020usu}.
This axiom says that there are no invisible lines in the bulk theory
(technically, that the category of lines has trivial center).
Violation of remote detectability signals multiple vacua
and therefore a decomposition,
much as cluster decomposition in other contexts \cite{Hellerman:2006zs}.

\subsection{Bulk-boundary map}

Let us now consider the bulk-boundary map between lines in the
three-dimensional bulk and on the two-dimensional boundary.
Let ${\cal C}$ be the category of lines which act trivially in the bulk.
Suppose we have a line in ${\cal C}$ which ends on the boundary,
defining an object in the two-dimensional
vertex operator algebra $V$.  We can describe this bulk-boundary
relation by a functor
\begin{equation}
F: \: {\cal C} \: \longrightarrow \: {\rm Rep}(V),
\end{equation}
(for Rep$(V)$ the category of representations of $V$)
which takes a line to the vector space of ways the line can end on the boundary,
giving point operators.

As observed in \cite[section 3.5]{Yu:2020twi},
a one-form symmetry that acts trivially in the bulk might act nontrivially
on the boundary, and the theory can still decompose, much as with Chan-Paton
factors and D-branes in two-dimensional theories.  Broadly speaking, the
different line operators in the three-dimensional bulk end on the various
two-dimensional sectors of the boundary theory.

A three-dimensional theory may have surface operators
which are not totally determined by the line operators.
In the case where the three-dimensional theory has only a local vacuum,
all the surfaces can built as condensations,
i.e.~networks of lines.
However, when there are multiple vacua,
as in the cases we are interested in, then this fails to be true.
The surfaces which are not built as a network of lines will end on a line
on the boundary.
These lines define the `action' of a trivially acting zero-form symmetry,
In two dimensions,
if one gauges a trivially-acting zero-form symmetry,
then one obtains an emergent global one-form symmetry
(and hence a decomposition).

From the decomposition conjecture~(\ref{eq:decomp-predict}),
the different universes and hence
the different ground states are labeled by elements of
the Pontryagin dual of the one-form symmetry group.
On the other hand, the surfaces in the bulk which enact a 2-form symmetry, come from gauging a trivially acting one-form symmetry.
So while the lines that the surface ends on has trivial action on the boundary, the surface itself is not necessarily trivial in the bulk.
This is summarized in the following diagram,
where $F$ is the functor that makes objects to the boundary:
\begin{equation}
\xymatrix{
\mbox{\underline{Bulk symmetry}} &
\mbox{\underline{Boundary symmetry}} 
\\
\mbox{trivially-acting one-form} \ar[r]^-{F}
\ar[d]^{\rm gauge}
& 
\mbox{trivially-acting zero-form} \ar[d]^{\rm gauge}
\\
\mbox{global two-form}
\ar[r]^-{F}
& \mbox{global one-form}
}
\end{equation}

\section{Examples}
\label{sect:exs}

In the next several subsections we will walk through examples
of the decomposition proposed in section~\ref{sect:decomp}.
Where possible, we will apply level-rank duality to perform
self-consistency tests.  In all cases, we will compare to the 
decomposition of the boundary
WZW model.  In particular, as reviewed
in section~\ref{sect:review-wzw}, decomposition is reasonably well-understood
in two-dimensional theories, and so we get solid consistency tests by
checking that the boundary WZW decomposition implied by the bulk
Chern-Simons decomposition matches existing two-dimensional results.

In each case, we will assume that levels are chosen so that the theories
are well-defined, but will not list those conditions explicitly.

\subsection{Chern-Simons$(SU(2))/B {\mathbb Z}_2$, $K=1$}
\label{sect:ex:so3}

In this section, we will reproduce a well-known result as a special
case of the decomposition prediction~(\ref{eq:decomp-predict}).

Specifically, we consider
gauging the $B {\mathbb Z}_2$ central one-form symmetry in
$SU(2)$ Chern-Simons theory.

Here, this $B {\mathbb Z}_2$ is not trivially-acting, and so no
decomposition is expected.  In particular, this gauging is known 
(see e.g.~\cite{Moore:1989yh})
to be equivalent to the $SO(3)$ Chern-Simons theory at the same level.
At the level of the path integral for the
gauge theory, this is discussed in appendix~\ref{sect:gauging-bk}.

We can understand this as a special case of the
decomposition prediction~(\ref{eq:decomp-predict}).
In the language of that statement, we identify
$A = {\mathbb Z}_2$, $H = SU(2)$, and $d: A \rightarrow H$ is the
inclusion map of the center, ${\mathbb Z}_2 \hookrightarrow \: 
SU(2)$.  Then, the kernel of $d$ vanishes, so $K = 1$, and
$G = H/A = SO(3)$.  This corresponds to the exact sequence
\begin{equation}
1 \: \longrightarrow \: 1 \: \longrightarrow \:
{\mathbb Z}_2 \: \stackrel{d}{\longrightarrow} \: SU(2)
\: \longrightarrow \: SO(3) \: \longrightarrow \: 1.
\end{equation}
Furthermore, in the case, since $K = 1$, the extension class
$[\omega] \in H^3(G,K)$ is trivial, $\omega = 1$, so $\phi^* \omega = 1$
and there is no discrete theta angle.

Putting this together, we see that the 
decomposition prediction~(\ref{eq:decomp-predict}) in this case is
\begin{equation}
\left[ \mbox{Chern-Simons}(SU(2)) / B {\mathbb Z}_2 \right] \: = \: 
\mbox{Chern-Simons}(SO(3)),
\end{equation}
which reproduces known results.

Let us also compute the line operator spectrum in this example.
This is a standard computation, but we will quickly outline it using
the tools of section~\ref{sect:lineops}, with an eye towards
later, more obscure, versions.  There are five
line operators in $SU(2)_4$
Chern-Simons, as listed in appendix~\ref{app:lineops}, which we denote
\begin{equation}
(0), \: (1), \: (2), \: (3), \: (4).
\end{equation}
We gauge a $B {\mathbb Z}_2$, with lines $\{ \ell_0, \ell_1 \}$,
where
\begin{equation}
\ell_i \times \ell_j \: = \: \ell_{i + j \mod 2},
\end{equation}
and which act on the $SU(2)_4$ lines as
\begin{equation}
\ell_0 \times L \: = \: L, \: \: \: \ell_1 \times L \: = \: (1) \times L,
\end{equation}
and with 
\begin{equation}
B(\ell_0, L) \: = \: +1, \: \: \:
B(\ell_1, 0) \: = \: B(\ell_1, 1) \: = \: B(\ell_1,4) \: = \: +1,
\: \: \:
B(\ell_1, 2) \: = \: B(\ell_1, 3) \: = \: -1.
\end{equation}
(Clearly, we can identify the action of this $B {\mathbb Z}_2$ with the
action of the lines $(0)$, $(1)$ in $SU(2)_4$.)
It is straightforward to check that this gives a well-defined action 
in the sense of section~\ref{sect:lineops}.  Applying the procedure there,
to get the lines of $SU(2)_4 / B {\mathbb Z}_2$,
\begin{itemize}
\item the lines (2), (3) are not
invariant under monodromies and so should be excluded,
\item from $(1) \times (1) = (0)$, the lines $(0)$ and $(1)$ should
be identified in the quotient, and 
\item from $(1) \times (4) = (4)$, the line $(4)$ is duplicated,
\end{itemize}
so that the $SU(2)_4/B {\mathbb Z}_2$ 
spectrum consists of the vacuum line and two
copies of $(4)$, which is the standard result for $SO(3)_4$.

Now, let us turn to the boundary theory.
On the boundary, this reduces to the statement
\begin{equation}
\left[ {\rm WZW}(SU(2)) / {\mathbb Z}_2 \right] \: = \: {\rm WZW}(SO(3)),
\end{equation}
which is standard.

\subsection{Chern-Simons$(SU(2)) \times [{\rm point}/B {\mathbb Z}_2]$, $K = 
{\mathbb Z}_2$}

Now, let us apply the decomposition prediction~(\ref{eq:decomp-predict})
to a different case, namely one in which we gauge a trivially-acting
$B {\mathbb Z}_2$ `acting' on an $SU(2)$ Chern-Simons theory,
uncoupled from the center one-form symmetry of the $SU(2)$ theory.

This is perhaps the cleanest example of a $B {\mathbb Z}_2$
gauging that acts trivially:  we gauge a $B {\mathbb Z}_2$ in bulk
that does nothing at all to the $SU(2)$.

Let us apply the decomposition prediction~(\ref{eq:decomp-predict}) to this
case.  Here, in the notation of~(\ref{eq:decomp-predict}), we take
$H = SU(2)$ and $A = {\mathbb Z}_2$; however, the map
$d: A \rightarrow H$ maps all of ${\mathbb Z}_2$ to $1$.
In this case, the kernel of $d$, $K$, is all of ${\mathbb Z}_2$,
and $G = H = SU(2)$.  The decomposition prediction for this case is
that
\begin{equation}
\left[ \mbox{Chern-Simons}(SU(2)) / B {\mathbb Z}_2 \right] \: = \:
\coprod_{\theta \in \hat{K}} \mbox{Chern-Simons}(SU(2)),
\end{equation}
two copies of the $SU(2)$ Chern-Simons theory.
Furthermore, in this case there are no nontrivial discrete theta angles,
hence the decomposition prediction can be written more simply as
\begin{equation}
\left[ \mbox{Chern-Simons}(SU(2)) / B {\mathbb Z}_2 \right] \: = \:
\coprod_2  \mbox{Chern-Simons}(SU(2)).
\end{equation}

Let us briefly consider the spectrum of line operators, following the
procedure discussed in section~\ref{sect:lineops}.
We describe the trivially-acting $B {\mathbb Z}_2$ in terms of
two lines $\{\ell_0, \ell_1\}$, where
\begin{equation}
\ell_i \times \ell_j \: = \: \ell_{i+j \mod 2},
\end{equation}
and with an action on the lines of $SU(2)_4$ given by
\begin{equation}
B(\ell_i, L) \: = \: +1, \: \: \:
\ell_i \times L \: = \: L.
\end{equation}
It is straightforward to check that this gives a well-defined action
in the sense of section~\ref{sect:lineops}.

Next, we compute the spectrum of $SU(2)_4 / B {\mathbb Z}_2$, for this
trivially-acting $B {\mathbb Z}_2$.
From the rules in section~\ref{sect:lineops},
\begin{itemize}
\item None of the original lines
of the $SU(2)$ Chern-Simons theory are omitted, as they all have trivial
monodromy under the generator $(a)$,
\item Since $(a) \times (a) = (0)$,
we see that in the gauged theory, $(a)$ and $(0)$
are identified with one
another, 
\item Since all of the original lines are invariant under fusion
($(a) \times (x) = (x)$), they are all duplicated.
\end{itemize}
As a result, the line operator spectrum of the gauged theory is two copies
of the line operator spectrum of the original $SU(2)$ Chern-Simons theory,
consistent with decomposition.  This result could also be obtained by
adding one new line $a$ to the lines of $SU(2)_4$, which interacts trivially
with all other lines, and then condensing $\{ (0), a \}$ in the ordinary
fashion, though as we discussed in section~\ref{sect:lineops}, it will
not always be possible to do that.

Next, we turn to the boundary theory.
In the boundary WZW model, bulk decomposition becomes the statement that
\begin{equation}
[ {\rm WZW}(SU(2)) / {\mathbb Z}_2 ] \: = \:
\coprod_2
{\rm WZW}(SU(2)).
\end{equation}
In the ${\mathbb Z}_2$ orbifold on the left, the
${\mathbb Z}_2$ acts trivially on the $SU(2)$ WZW model,
for which case ordinary two-dimensional decomposition predicts
exactly the statement above, that the completely-trivially-acting
${\mathbb Z}_2$ orbifold of a WZW model is just two copies of the
same WZW model.  Thus, the boundary theory matches results from
two-dimensional decomposition, as expected.

\subsection{Chern-Simons$(SU(2)) / B {\mathbb Z}_{4}$, $K = {\mathbb Z}_2$}
\label{sect:ex:su2-bz4}

Consider a $SU(2)$ Chern-Simons theory in three dimensions,
and gauge a $B {\mathbb Z}_4$ that acts via projecting to a 
$B{\mathbb Z}_2$ which acts as the center symmetry.
In this case, there is a trivially-acting $B {\mathbb Z}_2$,
so in broad brushstrokes one expects two copies of a $B {\mathbb Z}_2$-gauged
$SU(2)$ Chern-Simons theory.

Let us walk through the prediction of the decomposition
prediction~(\ref{eq:decomp-predict}) in this case.
Here, we have $H = SU(2)$ and $A = {\mathbb Z}_2$, with the map
$d: A \rightarrow SU(2)$ mapping the ${\mathbb Z}_4$ onto the
center ${\mathbb Z}_2$ of $SU(2)$.  Thus, the map $d$ is surjective,
but not injective:  its kernel $K = {\mathbb Z}_2$.
Similarly, 
\begin{equation}
G \: = \: H / {\rm im}\, d \: = \: SU(2)/ {\mathbb Z}_2 \: = \: SO(3).
\end{equation}

Putting this together, we see in this case that the
decomposition prediction~(\ref{eq:decomp-predict}) is
\begin{equation}   \label{eq:decomp:su2:z4}
\left[ \mbox{Chern-Simons}(SU(2)) / B {\mathbb Z}_4 \right] \: = \:
\mbox{Chern-Simons}(SO(3))_+ \: \coprod \: 
\mbox{Chern-Simons}(SO(3))_-,
\end{equation}
where the $\pm$ denote the two values of the discrete theta angle
coupling to the characteristic class defined by
$\beta_{\alpha}( w_G = w_{SO(3)})$, for $\alpha$ the class of the extension
\begin{equation} \label{eq:z2-bock}
1 \: \longrightarrow \:
{\mathbb Z}_2 \: \longrightarrow \: {\mathbb Z}_4 \: \longrightarrow \:
{\mathbb Z}_2 \: \longrightarrow \: 1,
\end{equation}
and where here, $w_{SO(3)} = w_2$, the second Stiefel-Whitney class.

Next, we will argue\footnote{
E.S. would like to thank Y.~Tachikawa for observing the pertinent
properties of $w_3$.}
that the characteristic class $\beta_{\alpha}(w_2)$
is the third Stiefel-Whitney class $w_3$.
From the Wu formula \cite[prob. 8-A]{ms} for Steenrod squares,
which map ${\rm Sq}^k: H^{\bullet}(X,{\mathbb Z}_2) \rightarrow
H^{\bullet+k}(X,{\mathbb Z}_2)$, $k \geq 0$:
\begin{equation}
{\rm Sq}^k( w_m(\xi) ) \: = \:
\sum_{t=0}^k \left( \begin{array}{c} k-m \\ t \end{array} \right)
w_{k-t}(\xi) \cup w_{m+t}(\xi)
\: = \: \sum_{t=0}^k
\left( \begin{array}{c} m-k+t-1 \\ t \end{array} \right)
w_{k-t}(\xi) \cup w_{m+t}(\xi),
\end{equation}
where each $w_j = w_j(\xi)$ for a real vector bundle $\xi$, and in the 
equality, we have used the fact that
\begin{eqnarray}
\left( \begin{array}{c} k-m \\ t \end{array} \right)
& = &
\frac{ (k-m)(k-m-1) \cdots (k-m-t+1) }{ t! },
\\
& = & 
(\pm) \frac{ (m-k)(m-k+1)\cdots(m-k+t-1)}{ t! }
\: \equiv \: \left( \begin{array}{c} m-k+t-1 \\ t \end{array} \right)
\mod 2.
\nonumber
\end{eqnarray}
(See e.g.~\cite{se1} for this and related observations.)
As a result, for any real vector bundle,
\begin{equation}
{\rm Sq}^1(w_2) \: = \: w_1 \cup w_2 \: + \: w_0 \cup w_3
\: = \: w_1 \cup w_2 \: + \: w_3,
\end{equation}
so if $w_1 = 0$, as is the case for $SO(3)$ bundles,
then $w_3 = {\rm Sq}^1(w_2)$.
(In principle, this is one explanation of why all $SO(3)$ bundles can be
constructed by twisting $SU(2)$ bundles by ${\mathbb Z}_2$ gerbes:  the gerbe
characteristic class determines not only the second Stiefel-Whitney class
$w_2$ of the $SO(3)$ bundles, but also $w_3$ via Sq$^1$, as above.)

Furthermore, the action of Sq$^1$ is the Bockstein homomorphism $\beta$
associated to the extension 
\begin{equation} \label{eq:z2z4}
1 \: \longrightarrow \: {\mathbb Z}_2 \: \longrightarrow \: {\mathbb Z}_4
\: \longrightarrow \: {\mathbb Z}_2 \: \longrightarrow \: 1,
\end{equation}
(see e.g.~\cite[section 4.L]{hatcher},)
meaning
\begin{equation}
{\rm Sq}^1(x) \: = \: \beta(x)  
\end{equation}
for any $x$.  The extension~(\ref{eq:z2z4}) above coincides with
$\alpha$ in the present case, so we see that in this example,
the discrete theta
angle 
couples to
\begin{equation}
\beta_{\alpha}(w_2) \: = \: 
{\rm Sq}^1(w_2),
\end{equation}
using~(\ref{eq:balpha-bock}).
We also see that in this example, this class
can be described even more simply as $w_3$,
the third Stiefel-Whitney class, as $w_3 = {\rm Sq}^1(w_2)$.

Now, 
on a three-manifold $M$, we can write Sq$^1(x)$ for any $x$ in terms of
the Wu class $\nu_1 \in H^1(M,{\mathbb Z}_2)$ as \cite[chapter 11]{ms}
\begin{equation}
{\rm Sq}^1(x) \: = \: \nu_1 \cup x.
\end{equation}
Furthermore, \cite[theorem 11.14]{ms}
\begin{equation}
\nu_1 \: = \: w_1(TM),
\end{equation}
so assembling these pieces, we have that
\begin{equation}
w_3(\xi) 
\: = \: {\rm Sq}^1(w_2(\xi)) 
\: = \: w_1(M) \cup w_2(\xi).
\end{equation}
As a result, the third Stiefel-Whitney class $w_3$ will only be
nontrivial on a nonorientable three-manifold $M$.
However, Chern-Simons theories are not defined on nonorientable spaces.

In section~\ref{sect:ex:u1-bzkn},
we will use level-rank duality to perform a self-consistency
check of decomposition in this case.

Now, let us check this prediction by computing
the line spectrum in this gauged Chern-Simons theory.
First, following section~\ref{sect:lineops}, we define a 
$B {\mathbb Z}_4$ by lines $\{ \ell_0, \ell_1, \ell_2, \ell_3 \}$
such that
\begin{equation}
\ell_i \times \ell_j \: = \: \ell_{i+j \mod 4},
\end{equation}
and which act on the lines of $SU(2)_4$ 
(described in appendix~\ref{app:lineops}) as follows:
\begin{equation}
B(\ell_{0,2}, L) \: = \: +1, \: \: \:
B(\ell_{1,3}, 0) \: = \: B(\ell_{1,3}, 1) \: = \:
B(\ell_{1,3}, 4) \: = \: +1, \: \: \:
B(\ell_{1,3}, 2) \: = \: B(\ell_{1,3}, 3) \: = \: -1,
\end{equation}
\begin{equation}
\ell_0 \times L \: = \: \ell_2 \times L \: = \:  L,
\: \: \:
\ell_1 \times L \: = \: \ell_3 \times L \: = \: (1) \times L.
\end{equation}
It is straightforward to check that this action of $B {\mathbb Z}_4$
on the lines of $SU(2)_4$ is well-defined in the sense of
section~\ref{sect:lineops}.
As $\ell_2$ acts trivially, this is also a non-effective action,
in the sense of section~\ref{sect:trivacting}.

Next, we follow the procedure outlined in section~\ref{sect:lineops}
to get the lines of $SU(2)_4 / B {\mathbb Z}_4$:
\begin{itemize}
\item Lines (2), (3) have $B(\ell_{1,3}, L) \neq +1$, and so are omitted.
\item Since $\ell_{1,3} \times (1) = (0)$, we identify the lines
$(0) \sim (1)$.
\item Since $\ell_i \times (4) = (4)$ for all $i$, we get four copies
of (4) in the spectrum of $SU(2)_4 / B {\mathbb Z}_4$,
and since $\ell_{0,2} \times (1) = (1)$, $\ell_{0,2} \times (0) = (0)$,
we get two copies of $(0) \sim (1)$.
\end{itemize}
Thus, we see that we get two copies of the lines of $SO(3)_4$,
consistent with expectations from decomposition.

Before going on, let us compute the lines in one more example,
specifically $SU(2)_4 / B {\mathbb Z}_{2p}$, where the
${\mathbb Z}_{2p}$ projects to the ${\mathbb Z}_2$ center of $SU(2)_4$,
with kernel ${\mathbb Z}_4$.  The lines of $B {\mathbb Z}_{2p}$ are 
$\{\ell_0, \cdots, \ell_{2p-1}\}$, where
\begin{equation}
\ell_i \times \ell_j \: = \: \ell_{i + j \mod 2p},
\end{equation}
and their action on $SU(2)_4$ is given by
\begin{equation}
B(\ell_{\rm even}, L) \: = \: +1, \: \: \:
B(\ell_{\rm odd}, 0) \: = \: +1 
\: = \: B(\ell_{\rm odd}, 1)
\: = \: B(\ell_{\rm odd}, 4),
\end{equation}
\begin{equation}
B(\ell_{\rm odd}, 2) \: = \: -1 \: = \: B(\ell_{\rm odd}, 3),
\end{equation}
\begin{equation}
\ell_{\rm even} \times L \: = \: L, \: \: \:
\ell_{\rm odd} \times L \: = \: (1) \times L.
\end{equation}
As before, it is straightforward to check that this action of
$B {\mathbb Z}_{2p}$ is well-defined in the sense of
section~\ref{sect:lineops}, and since $\{\ell_{\rm even}\}$ act
trivially, it is a non-effective action, in the sense of
section~\ref{sect:trivacting}.

Next, we follow the procedure outlined in section~\ref{sect:lineops}
to get the lines of $SU(2)_4 / B {\mathbb Z}_{2p}$:
\begin{itemize}
\item Lines (2), (3) have $B(\ell_{\rm odd}, L) \neq +1$, and so are omitted.
\item Since $\ell_{\rm odd} \times (1) = (0)$, we identify the lines
$(0) \sim (1)$.
\item Since $\ell_i \times (4) = (4)$ for all $i$, we get $2p$ copies of
(4), and since $\ell_{\rm even} \times (1) = (1)$, 
$\ell_{\rm even} \times (0) = (0)$, we get $p$ copies of $(0) \sim (1)$.
\end{itemize}
Altogether, we find $p$ copies of the lines of $SO(3)_4$, consistent with
expectations from decomposition, since $B {\mathbb Z}_p$ acts
trivially.

Before going on, let us briefly discuss the boundary theory.
The Chern-Simons decomposition~(\ref{eq:decomp:su2:z4}) becomes a
decomposition of WZW models, formally
\begin{equation}
\left[ {\rm WZW}(SU(2)) / {\mathbb Z}_4 \right] \: = \:
{\rm WZW}(SO(3))_+ \: \coprod \: 
{\rm WZW}(SO(3))_-.
\end{equation}
Here the ${\mathbb Z}_2$ discrete theta angle couples to
the image of the element of $H^3(B SO(3), {\mathbb Z}_2)$ 
(corresponding to third Stiefel-Whitney classes) in
$H^2( SO(3), {\mathbb Z}_2) = {\mathbb Z}_2$.
However, the generator of this group is Sq$^1(a)$, where $a$
generates $H^1(SO(3), {\mathbb Z}_2)$
and for reasons discussed previously, Sq$^1(a) = w_1(TM) \cup a$,
hence is nonzero only if the two-dimensional space is nonorientable.

We will consider various generalizations of this example,
returning to this example for special levels to utilize
level-rank duality consistency checks in
section~\ref{sect:ex:u1-bzkn}.

\subsection{Chern-Simons$(SU(n)) / B {\mathbb Z}_{np}$,
$K = {\mathbb Z}_{p}$}

Next, we will consider gauging the action of $B {\mathbb Z}_{np}$ on
$SU(n)$ Chern-Simons, where the ${\mathbb Z}_{np}$ acts by projecting
to the center ${\mathbb Z}_n$ of $SU(n)$, and study the
discrete theta angles for special values of
$n$ and $p$ beyond those discussed already.  

In terms of the
decomposition prediction~(\ref{eq:decomp-predict}), we take
$A = {\mathbb Z}_{np}$, $H = SU(n)$, and $d: A \rightarrow H$ acts
by projecting to $Z = {\mathbb Z}_n \subset Z(H)$.  
Then, the kernel $K = {\mathbb Z}_{p}$,
$G = SU(n)/{\mathbb Z}_n$, and we have the long exact sequence
\begin{equation}
1 \: \longrightarrow \: {\mathbb Z}_{\ell} \: \longrightarrow \:
{\mathbb Z}_{n\ell} \: \longrightarrow \: SU(n) \: \longrightarrow \:
SU(n)/{\mathbb Z}_{n} \: \longrightarrow \: 1.
\end{equation}
In general terms, decomposition~(\ref{eq:decomp-predict}) then predicts that
\begin{equation}
\left[ \mbox{Chern-Simons}(SU(n)) / BA \right]
\: = \:
\coprod_{\theta \in \hat{K}} \mbox{Chern-Simons}(SU(n)/{\mathbb Z}_n)_{\theta(\omega)},
\end{equation}
where the $\theta(\omega)$ are discrete theta angles coupling to the
characteristic class defined by $\beta_{\alpha}( w_{SU(n)/{\mathbb Z}_n})$,
where $w_{ SU(n)/{\mathbb Z}_n} \in H^2_{\rm sing}(B SU(n)/{\mathbb Z}_n, {\mathbb Z}_n)$ 
is a generalization
of the second Stiefel-Whitney class to $n \geq 2$, 
and $\beta_{\alpha}$ is the Bockstein map in the long exact sequence
associated to the extension
\begin{equation}
1 \: \longrightarrow \: K ( = {\mathbb Z}_p ) \: \longrightarrow \:
A (= {\mathbb Z}_{np}) \: \longrightarrow \:
Z ( = {\mathbb Z}_n ) \: \longrightarrow \: 1,
\end{equation}
with extension class $\alpha \in H^2_{\rm group}( Z, K)$.

We will evaluate this expression for some special cases in which we will
simplify the expression for discrete theta angles.
We will use \cite{bb}, which provides the cohomology of $SU(n)/{\mathbb Z}_n$,
which (modulo a degree shift) is essentially the same.  
(See also \cite{xgu,duan,kac,notbohm,km,borel}.)

First, consider the case that $p$ is a prime number that does not
divide $n$.  Then, from \cite[section 7]{bb}, 
\begin{equation}
H^{\bullet}_{\rm sing}(B SU(n)/{\mathbb Z}_n, {\mathbb Z}_p) \: = \:
H^{\bullet}_{\rm sing}(B SU(n), {\mathbb Z}_p),
\end{equation}
and so there is no ${\mathbb Z}_p$-valued characteristic class in degree three,
hence no discrete theta angle.  In this case, 
the decomposition above can be written more simply as
\begin{equation}
\left[ \mbox{Chern-Simons}(SU(n)) / BA \right]
\: = \:
\coprod_{p} \mbox{Chern-Simons}(SU(n)/{\mathbb Z}_n).
\end{equation}

Next, suppose that $p=2$, and $n=2m$ for $m$ odd.
From \cite[cor. 4.2]{bb}, the group $H^3_{\rm sing}( B SU(n)/{\mathbb Z}_n,
{\mathbb Z}_2) \neq 0$, and so for 
$w_{ SU(n)/{\mathbb Z}_n } \in H^2_{\rm sing}(B SU(n)/{\mathbb Z}_n,
{\mathbb Z}_n)$, 
we get a discrete theta angle coupling to $\beta_{\alpha}(w_{SU(n)/{\mathbb Z}_n})$,
the image of $w_{ SU(n)/{\mathbb Z}_n}$ 
under the Bockstein map associated to the extension
\begin{equation}
1 \: \longrightarrow \: {\mathbb Z}_p \: \longrightarrow \:
{\mathbb Z}_{pn} \: \longrightarrow \: {\mathbb Z}_n
\: \longrightarrow \: 1,
\end{equation}
with extension class 
$\alpha \in H^2_{\rm group}( {\mathbb Z}_n, {\mathbb Z}_p)$.
Since $p=2$, we can write $\beta_{\alpha}(w_{SU(n)/{\mathbb Z}_n}) 
= {\rm Sq}^1(w_{SU(n)/{\mathbb Z}_n})$,
as before, and also just as before,
it is only nonzero on nonoriented spaces, as we saw for 
the case of $SU(2)$ and $SO(3)$ theories in section~\ref{sect:ex:su2-bz4}.

Now, let us consider the corresponding boundary WZW model.
The bulk decomposition above predicts
\begin{equation}
\left[ {\rm WZW}( SU(n) ) / {\mathbb Z}_{n p} \right]
\: = \:
\coprod_{\theta \in \hat{K}} {\rm WZW}( SU(n)/{\mathbb Z}_n )_{\theta(\omega)}.
\end{equation}

Now, from ordinary two-dimensional decomposition, since there is
no discrete torsion in ${\mathbb Z}_n$,
\begin{equation}
\left[ {\rm WZW}( SU(n) ) / {\mathbb Z}_{n p} \right]
\: = \:
\coprod_{\ell} {\rm WZW}( SU(n) / {\mathbb Z}_n ).
\end{equation}
This is certainly consistent with the special cases computed above,
in which the bulk discrete theta angle vanishes.

\subsection{Chern-Simons$({\rm Spin}(n)) / B {\mathbb Z}_{2p}$, $K = 
{\mathbb Z}_p$}
\label{sect:cs:spin-n}

Next, we consider a simple generalization of the example above,
in which we gauge a $B {\mathbb Z}_{2p}$ action on Spin$(n)$ Chern-Simons,
in which the $B {\mathbb Z}_{2p}$ acts by first projecting to
$B {\mathbb Z}_2$ which acts through (a subgroup of) the center.
We begin by discussing the case that the ${\mathbb Z}_2$ is such that
Spin$(n)/{\mathbb Z}_2 = SO(n)$.  In the case that $n$ is divisible by four,
there is a second choice of ${\mathbb Z}_2$ subgroup, for which the
quotient Spin$(n)/{\mathbb Z}_2 \neq SO(n)$.
We will discuss the second case at the end of this section.

In terms of the decomposition prediction~(\ref{eq:decomp-predict}),
we take $A = {\mathbb Z}_{2p}$, $H = {\rm Spin}(n)$, and
$d: A \rightarrow H$ is the map that projects ${\mathbb Z}_{2p}$ onto
the ${\mathbb Z}_2$ in the center of Spin$(n)$ such that
Spin$(n)/{\mathbb Z}_2 = SO(n)$.  Then, the kernel of $d$ is
$K = {\mathbb Z}_p$, $G = H/A = SO(n)$, and we have the exact sequence
\begin{equation}
1 \: \longrightarrow \: {\mathbb Z}_p \: \longrightarrow \:
{\mathbb Z}_{2p} \: \longrightarrow \: {\rm Spin}(n) \: \longrightarrow \:
SO(n) \: \longrightarrow \: 1.
\end{equation}
This extension is nontrivial, and defines a discrete theta angle coupling
to $\beta_{\alpha}(w_{SO(n)})$, with $w_{SO(n)} = w_2$, the second
Stiefel-Whitney class, as before, and
the Bockstein homomorphism $\beta_{\alpha}$ is associated to
the extension
\begin{equation}
1 \: \longrightarrow \: {\mathbb Z}_p \: \longrightarrow \:
{\mathbb Z}_{2p} \: \longrightarrow \:
{\mathbb Z}_2 \: \longrightarrow \: 1
\end{equation}
of extension class $\alpha \in H^2_{\rm group}({\mathbb Z}_2,{\mathbb Z}_p)$.

Decomposition then predicts~(\ref{eq:decomp-predict})
\begin{equation} \label{eq:spinn-z2p}
\left[ \mbox{Chern-Simons}({\rm Spin}(n)) / B {\mathbb Z}_{2p} \right]
\: = \:
\coprod_{\theta \in \hat{\mathbb Z}_p}
\mbox{Chern-Simons}(SO(n))_{\theta},
\end{equation}
where the $\theta$ denotes the discrete theta angle coupling.

In the case that $p=2$, for the same reasons as discussed
in section~\ref{sect:ex:su2-bz4}, 
we can identify $\beta_{\alpha}(w_2)$ with $w_3$,
the third
Stiefel-Whitney class.
However, by the same reasoning as described
in subsection~\ref{sect:ex:su2-bz4}, the third Stiefel-Whitney class will
only be nontrivial on nonorientable three-manifolds.  Therefore,
on orientable three-manifolds, for $p=2$,
the statement of decomposition reduces to
\begin{equation}
\left[ \mbox{Chern-Simons}({\rm Spin}(n)) / B {\mathbb Z}_4 \right] \: = \:
\coprod_2 \mbox{Chern-Simons}(SO(n)).
\end{equation}

Next, let us briefly compare to the boundary WZW model.
On the boundary, from the decomposition~(\ref{eq:spinn-z2p}),
we have
\begin{equation}
\left[ {\rm WZW}( {\rm Spin}(n) ) / {\mathbb Z}_{2p} \right] \: = \:
\coprod_{\theta \in \hat{\mathbb Z}_p} {\rm WZW}( SO(n) )_{\theta},
\end{equation}

For the case $p=2$,
for the same reasons as noted in section~(\ref{sect:ex:su2-bz4}, 
for oriented spaces,
the discrete theta angles are trivial, as the characteristic class
they couple to vanish.
As a result, on oriented spaces, for $p=2$ we can equivalently write
\begin{equation}
\left[ {\rm WZW}( {\rm Spin}(n) ) / {\mathbb Z}_4 \right] \: = \:
\coprod_2 {\rm WZW}( SO(n) ).
\end{equation}

This is consistent with the prediction of decomposition in two dimensions
in this case.  As reviewed in section~\ref{sect:review-wzw},
essentially because there is no discrete torsion in a ${\mathbb Z}_2$
orbifold, in a two-dimensional WZW orbifold by ${\mathbb Z}_{2p}$
with trivially-acting ${\mathbb Z}_p$, we have
\begin{equation}
\left[ {\rm WZW}( {\rm Spin}(n) ) / {\mathbb Z}_4 \right] \: = \:
\coprod_p {\rm WZW}( SO(n) ).
\end{equation}
For $p=2$ this is certainly consistent with the bulk description.

So far we have discussed the case that the ${\mathbb Z}_{2p}$ maps to
${\mathbb Z}_2 \subset {\rm Spin}(n)$ such that 
\begin{equation}
{\rm Spin}(n)/{\mathbb Z}_2 \: =  \:
SO(n).  
\end{equation}
In the case that $n$ is divisible by four, there is another choice of 
${\mathbb Z}_2$ subgroup of the center of Spin$(n)$, which leads to
a quotient 
\begin{equation}
{\rm Spin}(n)/{\mathbb Z}_2 \: \neq \: SO(n),
\end{equation}
which for example
projects out the vector representation.  (See e.g.~\cite{Witten:1997bs} for
a discussion in a different context.)  This second quotient
group is sometimes denoted Semi-spin$(n)$, abbreviated $Ss(n)$
(see e.g.~\cite[section 11]{borel}).
Relevant material on the cohomology of $Ss(n)$ can be found in
e.g.~\cite[section 9]{bb}.

\subsection{Chern-Simons$({\rm Spin}(4n+2)) / B {\mathbb Z}_{4p}$,
$K = {\mathbb Z}_{p}$}

Let us consider the case of a Chern-Simons theory with gauge
group Spin$(4n+2)$ and a gauged $B {\mathbb Z}_{4p}$, where the
${\mathbb Z}_{4p}$ maps to the center (${\mathbb Z}_4$) of
Spin$(4n+2)$, with kernel $K = {\mathbb Z}_p$.

In terms of the decomposition prediction~(\ref{eq:decomp-predict}),
we take $A = {\mathbb Z}_{4p}$, $H = {\rm Spin}(4n+2)$, and $d: A \rightarrow H$
projects ${\mathbb Z}_{4p}$ onto the central
${\mathbb Z}_4 \subset {\rm Spin}(4n+2)$.  The kernel of $d$ is
$K = {\mathbb Z}_p$, $G = H / A = SO(4n+2)/{\mathbb Z}_2$, and we have
the exact sequence
\begin{equation}
1 \: \longrightarrow \: {\mathbb Z}_p \: \longrightarrow \:
{\mathbb Z}_{4p} \: \longrightarrow \:
{\rm Spin}(4n+2) \: \longrightarrow \: SO(4n+2)/{\mathbb Z}_2 \:
\longrightarrow \: 1.
\end{equation}
Decomposition then predicts~(\ref{eq:decomp-predict})
\begin{equation} \label{eq:spin4np2:bz4p}
\left[ \mbox{Chern-Simons}( {\rm Spin}(4n+2) ) / B {\mathbb Z}_{4p} \right]
\: = \:
\coprod_{\theta \in \hat{\mathbb Z}_p} 
\mbox{Chern-Simons}( SO(4n+2)/ {\mathbb Z}_2 )_{\theta(\omega)},
\end{equation}
where the discrete theta angle couples to a characteristic class
$\beta_{\alpha}(w_{{\rm Spin}(4n+2)/{\mathbb Z}_4})$ 
for $\beta_{\alpha}$ the Bockstein map associated
to the short exact sequence
\begin{equation}
1 \: \longrightarrow \: {\mathbb Z}_p \: \longrightarrow \:
{\mathbb Z}_{4p} \: \longrightarrow \: {\mathbb Z}_4 
\: \longrightarrow \: 1
\end{equation}
of extension class
$\alpha \in H^2_{\rm group}( {\mathbb Z}_4,
{\mathbb Z}_p)$.

Consider for example the case $p=2$.
From \cite[lemma 8.1]{bb}, $SO(4n+2)/{\mathbb Z}_2$ has one characteristic
class in $H^3( B SO(4n+2)/{\mathbb Z}_2, {\mathbb Z}_2 )$, related to
$w_3$ of a covering $SO(4n+2)$ bundle.

In the boundary WZW model, the decomposition~(\ref{eq:spin4np2:bz4p}) predicts
\begin{equation}
\left[ {\rm WZW}({\rm Spin}(4n+2)) /  {\mathbb Z}_{4p} \right]
\: = \:
\coprod_{\theta \in \hat{\mathbb Z}_p}
{\rm WZW}( SO(4n+2)/{\mathbb Z}_2)_{\theta}.
\end{equation}

Ordinary two-dimensional decomposition predicts in this case that
\begin{equation}
\left[ {\rm WZW}({\rm Spin}(4n+2)) /  {\mathbb Z}_{4p} \right]
\: = \:
\coprod_p {\rm WZW}( SO(4n+2)/{\mathbb Z}_2),
\end{equation}
essentially because there is no discrete torsion in a ${\mathbb Z}_4$
orbifold.

\subsection{Chern-Simons$({\rm Spin}(4n)) / B ({\mathbb Z}_2 \times {\mathbb Z}_{2p})$, $K = {\mathbb Z}_p$}
\label{sect:ex:cs-spin-4n}

Next, we consider the case of a $B ({\mathbb Z}_2 \times
{\mathbb Z}_{2p})$ action on
a Spin$(4n)$ Chern-Simons theory.  Here, Spin$(4n)$ has center
${\mathbb Z}_2 \times {\mathbb Z}_2$, and the
${\mathbb Z}_2 \times {\mathbb Z}_{2p}$ acts by first mapping to
the center.

In terms of the decomposition prediction~(\ref{eq:decomp-predict}),
we take $A = {\mathbb Z}_2 \times {\mathbb Z}_{2p}$,
$H = {\rm Spin}(4n)$, $d: A \rightarrow H$ maps $A$ onto the center,
$K = {\rm Ker}\, d = {\mathbb Z}_p$, hence we predict
\begin{equation} \label{eq:spin4n:z2-z2p}
\left[ \mbox{Chern-Simons}( {\rm Spin}(4n) ) / B ( {\mathbb Z}_2 \times
{\mathbb Z}_{2p}) \right] \: = \:
\coprod_{\theta \in \hat{\mathbb Z}_p}
\mbox{Chern-Simons}( SO(4n) / {\mathbb Z}_2 )_{\theta},
\end{equation}
where the discrete theta angle couples to $\beta_{\alpha}(w_{{\rm Spin}(4n)/{\mathbb Z}_2 \times {\mathbb Z}_2})$,
for $\beta_{\alpha}$ the Bockstein map associated to the
short exact sequence
\begin{equation}
1 \: \longrightarrow \: {\mathbb Z}_p \: \longrightarrow \:
{\mathbb Z}_2 \times {\mathbb Z}_{2p} \: \longrightarrow \:
{\mathbb Z}_2 \times {\mathbb Z}_2 \: \longrightarrow \:
1
\end{equation}
of extension class
$\alpha \in H^2_{\rm group}( {\mathbb Z}_2 \times {\mathbb Z}_2,
{\mathbb Z}_p)$.

Consider for example $p=2$.  From \cite[lemma 8.1]{bb},
$SO(4n)/{\mathbb Z}_2$ has one characteristic class in 
$H^3( B SO(4n)/{\mathbb Z}_2, {\mathbb Z}_2)$,
related to $w_3$ of a covering $SO(4n)$ bundle.

Now, let us consider this in the boundary WZW model.
The bulk decomposition~(\ref{eq:spin4n:z2-z2p}) predicts that
\begin{equation}   \label{eq:wzw:spin4n-z2z2p}
\left[ {\rm WZW}( {\rm Spin}(4n) ) / ( {\mathbb Z}_2 \times
{\mathbb Z}_{2p} ) \right] \: = \:
\coprod_{\theta \in \hat{\mathbb Z}_p} {\rm WZW}( SO(4n) / {\mathbb Z}_2 )_{\theta},
\end{equation}
where, as discussed in section~\ref{sect:boundary-wzw},
the boundary discrete theta angles $\theta$ correspond to choices
of discrete torsion, 
here in a $G = {\mathbb Z}_2 \times {\mathbb Z}_2$ orbifold.

We can understand those boundary discrete theta angles more precisely
by comparing
to the predictions of two-dimensional decomposition.
We have a $\Gamma = {\mathbb Z}_2 \times {\mathbb Z}_{2p}$ orbifold,
with trivially-acting $K = {\mathbb Z}_p$,
and $G = \Gamma/K = {\mathbb Z}_2 \times {\mathbb Z}_2$.
In principle, $G$ can contain discrete torsion, 
since $H^2({\mathbb Z}_2 \times {\mathbb Z}_2, U(1)) = {\mathbb Z}_2$,
so we should compute
to see if we get nontrivial discrete torsion in any factors.
Any such discrete torsion is the image of the extension class
in $H^2(G,K)$ corresponding to
\begin{equation}
1 \: \longrightarrow \: K \: \longrightarrow \: \Gamma \:
\longrightarrow \: G \: \longrightarrow \: 1
\end{equation}
under the map $K \rightarrow U(1)$ defined by the representation of $K$
corresponding to that universe, and the extension class is nontrivial;
nevertheless, as discussed in \cite[section 6.1]{Robbins:2020msp},
its image in $H^2(G,U(1))$ is trivial for both irreducible representations
of $K$.
As a result, two-dimensional decomposition predicts
\begin{equation}
\left[ {\rm WZW}( {\rm Spin}(4n) ) / ( {\mathbb Z}_2 \times
{\mathbb Z}_{2p} ) \right] \: = \:
\coprod_p {\rm WZW}( SO(4n) / {\mathbb Z}_2 ).
\end{equation}
In particular, the boundary discrete theta angles vanish.

In passing, we should observe that this is a nontrivial constraint.
The two choices of discrete torsion in the WZW model for
Spin$(4n)/{\mathbb Z}_2 \times {\mathbb Z}_2$ correspond to two
distinct quantum theories, each of which can be described as the
WZW model for $SO(4n)$, see
e.g.~\cite{Felder:1988sd,Gawedzki:2003pm,Gaberdiel:1995yk,Runkel:2008gr,Gawedzki:2009jj}.  Furthermore, in two dimensions, certainly there exist examples
in which both choices of discrete torsion appear.
For example, only slightly generalizing results in
\cite{Hellerman:2006zs}, 
\begin{eqnarray}
\left[ {\rm WZW}( {\rm Spin}(4n) ) / D_4 \right] & = &
{\rm WZW}( SO(4n) / {\mathbb Z}_2 )_+ \: \coprod \:
{\rm WZW}( SO(4n) / {\mathbb Z}_2 )_-,
\\
\left[ {\rm WZW}( {\rm Spin}(4n) ) / {\mathbb H} \right] & = &
{\rm WZW}( SO(4n) / {\mathbb Z}_2 )_+ \: \coprod \:
{\rm WZW}( SO(4n) / {\mathbb Z}_2 )_-,
\end{eqnarray}
where in both $D_4$ and ${\mathbb H}$ the ${\mathbb Z}_2$ center
is taken to act trivially, and the $\pm$ indicate the two choices of
discrete torsion.

However, because both the dihedral group $D_4$ and the group of unit
quaternions ${\mathbb H}$ are nonabelian, there is no Chern-Simons version
of the decompositions above.  That is fortuitous, as of the two
$SO(4n)/{\mathbb Z}_2$ WZW models, 
the one with nonzero discrete torsion also does not
have a Chern-Simons dual \cite{Gawedzki:2009jj,waldorf-priv}.

More generally, in order to get a two-dimensional decomposition
of $[$WZW$({\rm Spin}(4n))/\Gamma]$ to copies of 
WZW$(SO(4n)/{\mathbb Z}_2)$ 
with nontrivial discrete torsion, it is straightforward
to check that $\Gamma$ must be nonabelian, and so does not admit
a Chern-Simons description.

\subsection{Chern-Simons$(Sp(n))/B{\mathbb Z}_{2p}$, $K = {\mathbb Z}_p$}

Next, consider the case of a Chern-Simons theory with gauge group
$Sp(n)$ and a gauged $B {\mathbb Z}_{2p}$, where the ${\mathbb Z}_{2p}$
maps to the center (${\mathbb Z}_2$) of $Sp(n)$.

In terms of the decomposition prediction~(\ref{eq:decomp-predict}),
we take $A = {\mathbb Z}_{2p}$, $H = Sp(n)$, and $d: A \rightarrow H$
projects ${\mathbb Z}_{2p}$ onto the central
${\mathbb Z}_2 \subset Sp(n)$, with $K = {\rm Ker}\, d = {\mathbb Z}_p$.
Decomposition then predicts~(\ref{eq:decomp-predict})
\begin{equation}  \label{eq:spn:z2p}
\left[ \mbox{Chern-Simons}( Sp(n) ) / B {\mathbb Z}_{2p} \right]
\: = \:
\coprod_{\theta \in \hat{\mathbb Z}_p}
\mbox{Chern-Simons}(Sp(n) / {\mathbb Z}_2)_{\theta},
\end{equation}
where the discrete theta angle couples to a characteristic class
$\beta_{\alpha}(w_{Sp(n)/{\mathbb Z}_2})$ for $\beta_{\alpha}$ the Bockstein map associated
to the short exact sequence
\begin{equation}
1 \: \longrightarrow \: {\mathbb Z}_p \: \longrightarrow \: {\mathbb Z}_{2p}
\: \longrightarrow \: {\mathbb Z}_2 \: \longrightarrow \: 1
\end{equation}
of extension class
$\alpha \in H^2_{\rm group}({\mathbb Z}_2,{\mathbb Z}_p)$.
See e.g.~\cite[section 8]{bb} for results on pertinent characteristic
classes.

In the boundary WZW model, the bulk decomposition~(\ref{eq:spn:z2p} predicts
\begin{equation}
\left[ {\rm WZW}(Sp(n)) / {\mathbb Z}_2 \right]
\: = \:
\coprod_{\theta \in \hat{\mathbb Z}_p} 
{\rm WZW}( Sp(n)/{\mathbb Z}_2 )_{\theta}.
\end{equation}

Because there is no discrete torsion in a ${\mathbb Z}_2$ orbifold,
two-dimensional decomposition predicts in this case that
\begin{equation}
\left[ {\rm WZW}(Sp(n)) / {\mathbb Z}_2 \right]
\: = \:
\coprod_p {\rm WZW}( Sp(n)/{\mathbb Z}_2 ).
\end{equation}

\subsection{Chern-Simons$(U(1))_k / B {\mathbb Z}_{\ell p}$, $K = {\mathbb Z}_p$}
\label{sect:ex:u1-bzkn}

Consider a $U(1)_k$ Chern-Simons theory in three dimensions.
This theory has a global $B {\mathbb Z}_k$ symmetry which can be
gauged (see e.g.~\cite{Kreuzer:1993tf,Fuchs:1996dd},
\cite[appendix C]{Hsin:2018vcg}).  
It has slightly different properties depending upon whether $k$ is
even or odd (see e.g.~\cite[section 2.2]{Benini:2022hzx}):
\begin{itemize}
\item When $k$ is even, this theory has $k$ line operators, labelled by elements of
${\mathbb Z}_k$.  
If $k$ is 0 mod 8, 
then the $B {\mathbb Z}_k$
one-form symmetry generator has integer spin. If $k$ is 2 mod 8, then the one form generator has spin 1/4 and if $k$ is 4 mod 8, then the one-form symmetry generator is spin 1/2.
\item When $k$ is odd, the theory has $2k$ lines 
labelled by elements of ${\mathbb Z}_{2k}$ and is moreover a spin TQFT.  
The line with the label $k$ is the transparent fermion.
\end{itemize}

Now, consider gauging a $B {\mathbb Z}_{\ell p}$, where $\ell$ divides $n$,
where the ${\mathbb Z}_{\ell p}$ projects to ${\mathbb Z}_{\ell} \subset
{\mathbb Z}_k$, for 
that $B {\mathbb Z}_k$ above, with kernel $B {\mathbb Z}_p$.
Let us apply the decomposition prediction~(\ref{eq:decomp-predict}) to this
case.

In the language of~(\ref{eq:decomp-predict}), $A = {\mathbb Z}_{\ell p}$
and $H = U(1)$.  Here, the map $d: A \rightarrow H$ is given by
projecting $A = {\mathbb Z}_{\ell p}$ to a ${\mathbb Z}_{\ell} \subset 
{\mathbb Z}_k \subset U(1)$,
and so it has kernel $K = {\mathbb Z}_p$.  Furthermore, 
\begin{equation}
G \: = \: H/{\rm im}\, d \: = \: U(1) / {\mathbb Z}_{\ell} \: = \: U(1).
\end{equation}
In this case, $BU(1) = {\mathbb C}{\mathbb P}^{\infty}$ has no
odd degree cohomology, so there cannot be any discrete theta angles.
Thus, the decomposition prediction~(\ref{eq:decomp-predict}) for this case
is that
\begin{equation}  \label{eq:ex:u1}
\left[ \mbox{Chern-Simons}(U(1)_k) / B {\mathbb Z}_{\ell p} \right] \: = \:
\coprod_{p} \left[ \mbox{Chern-Simons}(U(1)_k)/B {\mathbb Z}_{\ell}
\right],
\end{equation}
a sum of $p$ theories 
(consistent with a trivially-acting $B {\mathbb Z}_p$)
with no discrete theta angles.

In particular, note that the right-hand side is a sum of 
$U(1)_k/B {\mathbb Z}_{\ell}$ Chern-Simons theories, which is not necessarily
the same as a union of $U(1)_k$ Chern-Simons theories.
Although as groups $U(1)/{\mathbb Z}_k = U(1)$, gauging a Chern-Simons
theory by a one-form symmetry is a bit different.  
For example, $U(1)_{4m}/B {\mathbb Z}_2 = U(1)_m$, from
\cite[section C.1]{Seiberg:2016rsg}.  (On the boundary, one has
a $U(1)$ WZW model, meaning a sigma model on $S^1$, with radius determined
by the level.  Gauging the $B {\mathbb Z}_k$ in bulk becomes gauging
a ${\mathbb Z}_k$ rotation in the boundary theory, which changes the
radius and hence the level.)

We can use level-rank duality to perform a consistency test.
Begin with the decomposition described in section~\ref{sect:ex:su2-bz4} at
level 1, namely,
\begin{equation}
\left[ \mbox{Chern-Simons}(SU(2)_1) / B {\mathbb Z}_4 \right] \: = \: 
\mbox{Chern-Simons}(SO(3)_1)_+ \: \coprod \:
\mbox{Chern-Simons}(SO(3)_1)_-.
\end{equation}
Here we have kept track of the discrete theta angle;
we only consider Chern-Simons theories on orientable manifolds,
so no discrete theta angle is visible, 
so the prediction of section~\ref{sect:ex:su2-bz4} in this case
is more simply
\begin{equation}
\left[ \mbox{Chern-Simons}(SU(2)_1) / B {\mathbb Z}_4 \right] \: = \: 
\coprod_2 \mbox{Chern-Simons}(SO(3)_1).
\end{equation}

From level-rank duality, we know \cite[sections 3.1, 3.2]{Hsin:2016blu}
\begin{equation}
U(1)_2 \: = \: U(1)_{-2} \: \leftrightarrow \: SU(2)_1,
\end{equation}
so we have that
\begin{equation}
\left[ \mbox{Chern-Simons}(U(1)_2) / B {\mathbb Z}_2 \right] \: = \:
\left[ SU(2)_1 / B {\mathbb Z}_2 \right] \: = \: 
\mbox{Chern-Simons}(SO(3)_1).
\end{equation}
Thus, we see from level-rank duality that
our decomposition in section~\ref{sect:ex:su2-bz4} implies
\begin{equation}
\left[ \mbox{Chern-Simons}(U(1)_2) / B{\mathbb Z}_4 \right] \: = \:
\coprod_2 \left[ \mbox{Chern-Simons}( U(1)_2) / B {\mathbb Z}_2 \right],
\end{equation}
which is a special case of the result~(\ref{eq:ex:u1}),
confirming in this case
that the decomposition prediction~(\ref{eq:decomp-predict})
is giving results compatible with this example of level-rank duality.

Next, we compute the spectrum of line operators in
$U(1)_8 / B {\mathbb Z}_{2p}$, using the methods of
section~\ref{sect:lineops}, hwere in the gauging,
the ${\mathbb Z}_{2p}$ projects to ${\mathbb Z}_2$ with
trivially acting ${\mathbb Z}_p$.  We describe the
${\mathbb Z}_{2p}$ by a set of lines
$\{ \ell_i \}$, $i \in \{0, \cdots, 2p-1\}$, where
\begin{equation}
\ell_i \times \ell_j \: = \: \ell_{i+j \mod 8}.
\end{equation}
$U(1)_8$ has eight lines, labelled
\begin{equation}
(0), \: (1), \: (2), \: (3), \: (4), \: (5), \: (6), \:
(7)
\end{equation}
whose properties are listed in appendix~\ref{app:lineops},
and for which $\{(0), (4)\}$ encode a $B {\mathbb Z}_2$.
The action of $B {\mathbb Z}_{2p}$ on the lines of $U(1)_8$ is given
as follows:
\begin{equation}
B(\ell_{\rm even}, L) \: = \: +1, \: \: \:
B(\ell_{\rm odd}, L) \: = \: B(4,L),
\end{equation}
\begin{equation}
\ell_{\rm even} \times L \: = \: L, \: \: \:
\ell_{\rm odd} \times L \: = \: (4) \times L,
\end{equation}
using the monodromies and fusion algebra described in
appendix~\ref{app:lineops}.  It is straightforward that this gives
a well-defined action in the sense of section~\ref{sect:lineops}.

Next, we compute the spectrum of lines in $U(1)_8 / B {\mathbb Z}_8$,
following the procedure of section~\ref{sect:lineops}.
\begin{itemize}
\item The lines (1), (3), (5), (7) have $B(\ell_{\rm odd}, L) = -1 \neq +1$,
and so are excluded.
\item $\ell_1 \times (0) = (4)$, $\ell_1 \times (2) = (6)$,
so we identify $(0) \sim (4)$, $(2) \sim (6)$.
\item $\ell_{\rm even} \times L = L$, so we get $p$ copies of
$(0) \sim (4)$ and $(2) \sim (6)$.
\end{itemize}
Thus, the resulting spectrum is $p$ copies of
$\{ (0) \sim (4), (2) \sim (6) \}$, which is the same
as $p$ copies of the line operator spectrum of
$U(1)_8 / B {\mathbb Z}_2$, as expected from decomposition,
since there is a trivially-acting ${\mathbb Z}_p$.

Next, let us compare to boundary WZW models. 
A (boundary) WZW model for the group $U(1)$ is the same
as a $c=1$ free scalar, of radius determined by the level.
(See e.g.~\cite[appendix C.1]{Seiberg:2016rsg} for discussions of
the RCFTs arising at particular values of the level.)
Gauging the bulk one-form symmetry corresponds to orbifolding the
boundary $c=1$ theory, which just changes the radius of the target-space
circle
in that boundary $c=1$ theory.

In a two-dimensional sigma model with target $S^1$, if we orbifold
by a ${\mathbb Z}_{kp}$ where ${\mathbb Z}_p \subset {\mathbb Z}_{kp}$
acts trivially, then from two-dimensional decomposition, the resulting
theory is equivalent to $p$ copies of the effectively-acting
${\mathbb Z}_k$ orbifold, precisely matching~(\ref{eq:ex:u1}),
as expected.

\subsection{Exceptional groups}

So far we have discussed quotients of Chern-Simons theories for
the gauge groups $SU(n)$, Spin$(n)$, and $Sp(n)$.
We can also consider cases with exceptional gauge groups.
Although $G_2$, $F_4$, and $E_8$ have no center,
the group $E_6$ has center ${\mathbb Z}_3$, and $E_7$ has
center ${\mathbb Z}_2$ (see e.g.~\cite[appendix A]{Distler:2007av}).

For example, applying decomposition~(\ref{eq:decomp-predict}),
for a ${\mathbb Z}_{3p}$ that acts on $E_6$ by projecting to
the ${\mathbb Z}_3$ center with kernel ${\mathbb Z}_p$,
\begin{equation}
\left[ \mbox{Chern-Simons}(E_6)/B {\mathbb Z}_{3p} \right]
\: = \: 
\coprod_{\theta \in \hat{\mathbb Z}_p}
 \mbox{Chern-Simons}( E_6/{\mathbb Z}_3 )_{\theta},
\end{equation}
where the discrete theta angle couples to $\beta_{\alpha}(w_{E_6/{\mathbb Z}_3})$, for
$\beta_{\alpha}$ the Bockstein map associated to the short exact sequence
\begin{equation}
1 \: \longrightarrow \: {\mathbb Z}_p \: \longrightarrow \:
{\mathbb Z}_{3p} \: \longrightarrow \:
{\mathbb Z}_{3} \: \longrightarrow \: 1
\end{equation}
of extension class
$\alpha \in H^2_{\rm group}( {\mathbb Z}_3, {\mathbb Z}_p )$.

Similarly, from decomposition~(\ref{eq:decomp-predict}),
for a ${\mathbb Z}_{2p}$ that acts on $E_7$ by projecting to the
${\mathbb Z}_2$ center with kernel ${\mathbb Z}_p$,
\begin{equation}
\left[ \mbox{Chern-Simons}(E_7)/B {\mathbb Z}_{2p} \right]
\: = \: 
\coprod_{\theta \in \hat{\mathbb Z}_p}
 \mbox{Chern-Simons}( E_7/{\mathbb Z}_2 )_{\theta},
\end{equation}
where the discrete theta angle couples to $\beta_{\alpha}(w_{E_7/{\mathbb Z}_2})$, for
$\beta_{\alpha}$ the Bockstein map associated to the short exact sequence
\begin{equation}
1 \: \longrightarrow \: {\mathbb Z}_p \: \longrightarrow \:
{\mathbb Z}_{2p} \: \longrightarrow \:
{\mathbb Z}_{3} \: \longrightarrow \: 1
\end{equation}
of extension class
$\alpha \in H^2_{\rm group}( {\mathbb Z}_2, {\mathbb Z}_p )$.

In both cases, in the boundary WZW model, 
this reduces to two-dimensional decomposition of a WZW orbifold,
with the discrete theta angles becoming choices of discrete torsion.
In both cases, as the orbifolds involve cyclic groups,
discrete torsion is trivial, so the boundary decomposition
yields just a disjoint union of copies of the same WZW orbifold.

\subsection{Chern-Simons$(H_1 \times H_2) / BA$}

For completeness, let us also briefly discuss decomposition in gauged
Chern-Simons theories whose gauge groups are a product of
Lie groups.  
Specifically, consider the gauge of a gauged $BA$ action,
for $A$ finite and abelian, on a Chern-Simons theory for
$H_1 \times H_2$ (at various levels, such that the gauge theory is
well-defined on the given three-manifold).  Bulk decomposition
takes the same form as~(\ref{eq:decomp-predict}):
\begin{equation}
\left[ {\rm Chern-Simons}(H_1 \times H_2) / BA \right]
\: = \: \coprod_{\theta \in \hat{K}} {\rm Chern-Simons}(G)_{\theta},
\end{equation}
where
\begin{equation}
1 \: \longrightarrow \: K \: \longrightarrow \: 
A \: \stackrel{d}{\longrightarrow} \: H_1 \times H_2 \:
\longrightarrow \: G \: \longrightarrow \: 1,
\end{equation}
and the discrete theta angle couples to $\beta_{\alpha}(w_G)$,
for $\beta_{\alpha}$ the Bockstein homomorphism associated to
\begin{equation}
1 \: \longrightarrow \: K \: \longrightarrow \: A \:
\longrightarrow \: Z \: \longrightarrow \: 1,
\end{equation}
classified by $\alpha \in H^2_{\rm group}(Z, K)$,
where $Z$ is a subgroup of the product of the centers of $H_{1,2}$,
given by
the image of $d$.

On the boundary, as before, this reduces to decomposition in the
two-dimensional theory, here
\begin{equation}
\left[ {\rm WZW}(H_1 \times H_2) / A \right] \: = \: 
\coprod_{\theta \in \hat{K}} {\rm WZW}(G)_{\theta},
\end{equation}
where the discrete theta angles $\theta$ now correspond to choices of
discrete torsion in a 
\begin{equation}
[{\rm WZW}(H_1 \times H_2) / Z]
\end{equation}
orbifold.  Essentially because $A$ is abelian,
for ultimately the same reasons as in
section~\ref{sect:ex:cs-spin-4n}, the discrete torsion is trivial on each
universe.

\subsection{Finite 2-group orbifolds}

So far we have focused on Chern-Simons theories in three dimensions,
but the same ideas apply to the finite 2-group orbifolds discussed
in \cite{Pantev:2022kpl}.  There, orbifolds by 2-groups $\Gamma$ were described,
where $\Gamma$ is an extension
\begin{equation}
1 \: \longrightarrow \: BK \: \longrightarrow \: \Gamma \: \longrightarrow
\: G \: \longrightarrow \: 1,
\end{equation}
where $G$, $K$ are both finite and $K$ is abelian, determined by
$[\omega] \in H^3_{\rm group}(G,K)$.  
Now, $\Gamma$ can also be described by a crossed
module $\{d: A \rightarrow H\}$, corresponding to a four-term exact sequence
of ordinary groups
\begin{equation}
1 \: \longrightarrow \: K \: \longrightarrow \: A \:
\stackrel{d}{\longrightarrow} \: H \: \longrightarrow \: G 
\: \longrightarrow \: 1,
\end{equation}
also determined (up to equivalences) by $[\omega] \in H^3_{\rm group}(G,K)$
(see e.g.~\cite[section IV.9]{hs} for related observations).

In this language, we can write the 2-group orbifold $[X/\Gamma]$ in terms
of the crossed module as
\begin{equation}
[X/\Gamma] \: = \: \left[ [X/H] / BA \right],
\end{equation}
at least for a presentation in which $A$ is abelian.

For this slightly different physical realization in terms of finite groups,
the statement of decomposition~(\ref{eq:decomp-predict}) is modified, but
only slightly:
\begin{equation}
[X/\Gamma] \: = \: \left[ [X/H] / BA \right]
\: = \: \oplus_{\theta \in \hat{K}} [X/G]_{\omega(\theta)},
\end{equation}
where the discrete torsion (formerly discrete theta angle)
$\omega(\theta)$ is defined by $\phi^* \omega$.
In this sense, the decomposition described in this paper is simply
a variation on the 2-group orbifold decomposition described in
\cite{Pantev:2022kpl}.  The fact that bulk discrete theta angles
(here, $C$-field analogues of discrete torsion) become (ordinary)
discrete torsion in the boundary theory was also observed in
\cite[section 3.2]{Pantev:2022kpl}.

In passing, we should also observe that results in finite 2-group orbifolds
have a qualitatively different form.  For example,
\cite[section 4.4]{Pantev:2022kpl} described an orbifold
by a 2-group extension
\begin{equation} \label{eq:2gp-ext}
1 \: \longrightarrow \: B {\mathbb Z}_2 \: \longrightarrow \: \Gamma
\: \longrightarrow \: ( {\mathbb Z}_2 )^3 \: \longrightarrow \: 1.
\end{equation}
In this case, $[X/\Gamma]$ is equivalent to a pair of copies of
$[X/ ({\mathbb Z}_2)^3]$ orbifolds, each with a different $C$ field
discrete torsion in $H^3_{\rm group}( ({\mathbb Z}_2)^3, U(1))$, which is nontrivial
even on $T^3$.
One could imagine an analogous theory here, such as
a quotient of $SU(2)^3$ Chern-Simons by $BA$ (for $A$ a finite abelian
group, with $K = {\mathbb Z}_2$ kernel, say) 
that leads to a disjoint union of $SO(3)^3$ Chern-Simons theories.
Here, however, in the case of Chern-Simons theories, no analogue of
$C$ field discrete torsion is present for $T^3$, partly because
(as noted in section~\ref{sect:nontriv}) 
the pertinent Bockstein homomorphism vanishes.
Part of the difference between these two theories is that in the Chern-Simons
case, the pertinent exact sequence of finite groups has the form
\begin{equation} \label{eq:su23-se}
1 \: \longrightarrow \: {\mathbb Z}_2 \: \longrightarrow \:
A \: \longrightarrow \: ( {\mathbb Z}_2 )^3 \: \longrightarrow \: 1,
\end{equation}
whereas by contrast the analogous sequence in~\cite{Pantev:2022kpl},
namely~(\ref{eq:2gp-ext}), can be alternately encoded
as a four-term sequence
\begin{equation}
1 \: \longrightarrow \: {\mathbb Z}_2 \: \longrightarrow \: 
P' \: \longrightarrow \: Q' \: \longrightarrow \: ( {\mathbb Z}_2 )^3
\: \longrightarrow \: 1,
\end{equation}
which realizes an element of $H^3_{\rm group}( ({\mathbb Z}_2)^3, {\mathbb Z}_2)$.
By contrast, the short exact sequence~(\ref{eq:su23-se}) realizes
an element of $H^2_{\rm group}( ({\mathbb Z}_2)^3, {\mathbb Z}_2)$, 
cohomology of different degree; the crossed module construction realizes 
a 2-group, but involves different groups.

\section{Boundary $G/G$ models}
\label{sect:boundary-g-g}

For completeness, in this section we include a different example
of a decomposition.

Consider gauged WZW models $G/H$ at level $k$, on the boundary of
a three-dimensional theory.  Because the $H$ action being gauged
is an adjoint action \cite{Witten:1991mk}, if the center $Z(H)$
of $H$ is nonzero, it acts trivially, and in two dimensions, the
resulting gauged WZW model decomposes into universes indexed by
irreducible representations of $Z(H)$.

Now, let us compare to the bulk theory.
From \cite[section 3]{Moore:1989yh},
for the gauged WZW model $G/H$ at level $k$, the bulk three-dimensional
theory is a $(G \times H)/Z$ gauge theory, with $Z$ the commmon center of
$G$ and $H$,
with action
\begin{equation} \label{eq:bulk-gauged-wzw}
k \ell S_{\rm CS}(G) \: - \: k  S_{\rm CS}(H),
\end{equation}
where $\ell$ is the index of the embedding $H \hookrightarrow G$,

Consider the special case of
the two-dimensional $G/G$ model, on the boundary of a 
three-dimensional theory.  
The $G/G$ model decomposes into
universes indexed by the integrable representations.
(In principle this is because it is a unitary topological
field theory \cite{Durhuus:1993cq,Moore:2006dw}; 
the specific relation to decomposition is
via noninvertible symmetries, as discussed in 
\cite{Komargodski:2020mxz,Huang:2021zvu}.)
From the discussion above, the bulk dual to the boundary $G/G$ model
appears to
have an identically-zero action~(\ref{eq:bulk-gauged-wzw}).
Since the boundary theory is a topological field theory,
this would be trivially consistent.

For more general boundary $G/H$ gauged WZW models,
the bulk action~(\ref{eq:bulk-gauged-wzw}) does not vanish identically.
Decomposition of the boundary suggests that the bulk may also 
decompose, in which case the bulk theory should admit a 
global two-form symmetry.  We leave elucidating that symmetry for
future work.

\section{Conclusions}

In this paper, we have discussed decomposition in three-dimensional
Chern-Simons theories with gauged noneffectively-acting one-form
symmetries.  In the bulk decomposition, the different universes
of the decomposition have discrete theta angles coupling to
bundle characteristic classes, specifically, images under Bockstein
maps of canonical degree-two characteristic classes.  On the boundary,
those map to choices of discrete torsion, and the bulk decomposition
becomes a standard orbifold decomposition, involving WZW models, which
serves as a strong consistency test.  

There are many directions this work could be taken.
One example would be to consider decomposition in
gauged Chern-Simons theories
in which the original theory has a discrete theta angle,
analogous to decomposition in two-dimensional orbifolds with
discrete torsion \cite{Robbins:2020msp}.
Another example would be to consider decomposition in Chern-Simons-matter
theories, rather than pure Chern-Simons.
Similarly, it would be interesting to consider decomposition in
holomorphic Chern-Simons \cite{rtthesis}, or deformations of
Chern-Simons theories, as arise when studying disk instanton corrections
in string compactifications.

It would also be interesting to understand dimensional reduction of
decomposition to two dimensions.  The dimensional reduction of pure
Chern-Simons is the two-dimensional $G/G$ model (which as a unitary TFT
already admits a decomposition 
\cite{Durhuus:1993cq,Moore:2006dw,Komargodski:2020mxz,Huang:2021zvu}), 
and the $BK$ symmetry
in three dimensions should become a $K \times BK$ symmetry in the 
two-dimensional theory.

In condensed matter physics, there exists a realization of
Chern-Simons theories known as the Levin-Wen model  \cite{Levin:2004mi},
and it would be interesting to consider this story in that setting.

In a different direction, Chern-Simons theories can also arise
on boundaries of four-dimensional theories, and it would be interesting
to study decomposition in that context, perhaps relating it to the
decomposition arising after instanton restriction in
\cite{Tanizaki:2019rbk}.  There, the instanton restriction resulted
in a disjoint union of four-dimensional Yang-Mills theories with
theta angle terms of the form
\begin{equation}
\frac{1}{8 \pi^2} \frac{2 \pi m}{k} \int {\rm Tr}\,F\wedge F,
\end{equation}
for $m \in \{0, 1, \cdots, k-1\}$, which implements the
restriction on instantons.  On a boundary, that would become
a disjoint union of theories, whose actions have Chern-Simons terms of the
form
\begin{equation}
\frac{1}{8 \pi^2} \frac{2 \pi m}{k} \int \omega_{\rm CS},
\end{equation}
clearly related to the disjoint unions of Chern-Simons theories we discuss
in this paper.  We leave such considerations for future work.

\section{Acknowledgements}

We would like to thank C.~Closset,
S.~Datta, J.~Distler, D.~Berwick~Evans, T.~Gomez,
S.~Gukov, L.~Lin,
D.~Robbins, I.~Runkel, U.~Schreiber, J.~Stasheff,
Y.~Tachikawa, T.~Vandermeuelen, and K.~Waldorf for useful
discussions.
We would further like to thank M.~Yu for initial collaboration and many
discussions.
T.P. was partially supported by NSF/BSF grant DMS-2200914,
NSF grant DMS-1901876, and Simons Collaboration grant number 347070.
E.S. was partially supported by NSF grant
PHY-2014086.

\appendix

\section{Line operators}
\label{app:lineops}

In this appendix we briefly review some basics of line operators
in Chern-Simons theories and their quantum numbers, to make this
paper self-contained.

In general, the line operators in a Chern-Simons theory at level $k$
correspond to integrable representations, which for
a model at level $k$, are the representations of highest weight
$\lambda$ satisfying the unitarity bound
\cite[equ'n (9.30)]{Ginsparg:1988ui}
\begin{equation}
2 \frac{ \psi \cdot \lambda}{\psi^2} \: \leq \: k,
\end{equation} 
for $\psi$ the highest weight of the adjoint representation.
(For example, for $SU(n)$ the integrable representations at any
level are classified by Young diagrams of width bounded by the level.)
Similarly, for a given WZW primary associated to an integrable representation
of highest weight $\lambda$, the $L_0$ eigenvalue is
\cite[equ'n (15.87)]{DiFrancesco:1997nk}
\begin{equation}
h \: = \: \frac{ (\lambda, \lambda + 2 \rho) }{ 2(k+g) },
\end{equation}
where $g$ is the dual Coxeter number and $\rho$ the Weyl vector
(half-sum of positive roots).
In passing, a representation is integrable if and only if its dual is
integrable, and it and its dual define WZW primaries of the same
conformal weight, see e.g.~\cite[section 8.3]{Distler:2007av}.
Similarly, the quantum dimension is given by
\cite[equ'n (16.66)]{DiFrancesco:1997nk}
\begin{equation}
\prod_{\alpha > 0} \frac{
\sin\left( \frac{ \pi( \lambda + \rho, \alpha) }{ k + g} \right)
}{
\sin\left( \frac{ \pi(\rho,\alpha) }{ k+g} \right)
}.
\end{equation}

For use in examples in the text,
the line operators of $SU(2)_4$ are\footnote{
We would like to thank M.~Yu for providing the results for line operators
of $SU(2)_4$ and $U(1)_8$ listed in this appendix.
}
\begin{align}
    \begin{array}{c|cccc}
    SU(2)_4 & \text{Integrable rep.} & \tilde{\lambda} &h & \text{q-dim}\\\hline
         (0)& ${\bf 1}$ & [0,4] & 0 & 1 \notag \\
         (1)& $\tiny\yng(4)$ & [4,0] & 1 & 1  \notag\\
         (2)& $\tiny\yng(1)$ & [1,3] & 1/8 & \sqrt{3} \notag\\
         (3)& $\tiny\yng(3)$ & [3,1] & 5/8 & \sqrt{3} \notag \\
         (4)& $\tiny\yng(2)$ & [2,2] &1/3 & 2,
    \end{array}
\end{align}
where $\tilde{\lambda}$ denotes the Dynkin label of each line, $h$ is the
conformal weight of the corresponding boundary chiral primary as above,
and q-dim denotes the quantum dimension.

The fusion algebra of $SU(2)_4$ lines can be computed with 
the program Kac \cite{kac1},
and that algebra is given below:
\begin{center}
\begin{tabular}{ccl}
$(0) \times (0) = (0)$, & \qquad & $(2) \times (2) = (0) + (4)$,
\\
$(0) \times (1) = (1)$, & & $(2) \times (3) = (1) + (4)$,
\\
$(0) \times (2) = (2)$, & & $(2) \times (4) = (2) + (3)$,
\\
$(0) \times (3) = (3)$, & & $(3) \times (3) = (0) + (4)$,
\\
$(0) \times (4) = (4)$, & & $(3) \times (4) = (2) + (3)$,
\\
$(1) \times (1) = (0)$, & & $(4) \times (4) = (0) + (1) + (4)$.
\\
$(1) \times (2) = (3)$,
& & \\
$(1) \times (3) = (2)$,
& & \\
$(1) \times (4) = (4)$,
& &
\end{tabular}
\end{center}
We see that the lines $(0)$, $(1)$ are mutually transparent, and their
fusion products have the structure of the group ${\mathbb Z}_2$.

From the table above, it is straightforward to compute the monodromies
of the line $(1)$ about other lines, using
\begin{equation}
B(a,b) \: = \: \exp\left( 2 \pi i \left( h(a \times b) - h(a) - h(b) \right)
\right),
\end{equation}
and one finds
\begin{eqnarray}
B(1,1) & = & +1,
\\
B(1,2) & = & -1,
\\
B(1,3) & = & -1, 
\\
B(1,4) & = & +1,
\end{eqnarray}
so that all monodromies are in $\{ \pm 1 \}$, as expected for a
$B {\mathbb Z}_2$, and also consistent with the fact that (2) and (3)
correspond to Wilson lines for an odd number of copies of
the ${\mathbb 2}$ representation.

Similarly, it will be useful later to write down the fusion
algebra for $U(1)_8$.  Here, there are eight lines, labelled $(0)$ through
$(7)$, 
with conformal weights and quantum dimensions
\begin{center}
\begin{tabular}{c|cc}
$U(1)_8$ & h & q-dim \\ \hline
$(0)$ & $0$ & $1$ \\
$(1)$ & $1/16$ & $1$ \\
$(2)$ & $1/4$ & $1$ \\
$(3)$ & $9/16$ & $1$ \\
$(4)$ & $1$ & $1$ \\
$(5)$ & $9/16$ & $1$ \\
$(6)$ & $1/4$ & $1$ \\
$(7)$ & $1/16$ & $1$
\end{tabular}
\end{center}
and the fusion algebra acts by addition, as
\begin{equation}
(a) \times (b) \: = \: (a + b \mod 8).
\end{equation}

From the table of lines above, it is clear that there is a 
$B {\mathbb Z}_2$ corresponding to the lines $\{(0), (4) \}$.
For use in section~\ref{sect:ex:u1-bzkn}, we list here pertinent
monodromies:
\begin{equation}
B((0), L) \: = \: +1,
\: \: \: B(4,0) \: = \:
B(4,2) \: = \: B(4,4) \: = \: B(4,6) \: = \: +1,
\end{equation}
\begin{equation}
B(4,1) \: = \: B(4,3) \: = \: B(4,5) \: = \: B(4,7) \: = \: -1.
\end{equation}

\section{Overview of crossed modules}
\label{app:crossed-module}

In this paper we have described 2-groups using crossed modules.
As they play an important role in the decomposition statment in
three-dimensional Chern-Simons theories, to make this paper self-contained
we include a brief overview here.

Briefly, a crossed module consists of the following data:
\begin{itemize}
\item a pair of groups $G_0$, $G_1$,
\item a group homomorphism $d: G_1 \rightarrow G_0$,
\item a group homomorphism $\alpha: G_0 \rightarrow {\rm Aut}(G_1)$,
\end{itemize}
such that 
\begin{enumerate}
\item the composition 
\begin{equation}
G_1 \: \stackrel{d}{\longrightarrow} \: G_1 \:
\stackrel{\alpha}{\longrightarrow} \: {\rm Aut}(G_1)
\end{equation}
is the conjugation action of $G_1$ on itself, meaning
\begin{equation}
\alpha\left( d(g_1) \right)(h) \: = \: g_1 h g_1^{-1},
\end{equation}
for $g_1, h \in G_1$, or equivalently that 
\begin{equation}
\xymatrix{
G_1 \times G_1 \ar[rr]^{d \times {\rm Id}} \ar[dr]_{{\rm Ad}}
& & G_0 \times G_1
\ar[dl]^{\alpha} 
\\
& G_1 &
}
\end{equation}
commutes,
\item $d$ is equivariant for the $G_0$ action on the source and target,
meaning
\begin{equation}
d\left( \alpha(g_0)(h) \right) \: = \: g_0 d(h) g_0^{-1}
\end{equation}
for $g_0, h \in G_0$, or equivalently that
\begin{equation}
\xymatrix{
G_0 \times G_1 \ar[r]^-{\alpha} \ar[d]_{{\rm Id}\times d} & G_1 \ar[d]^d
\\
G_0 \times G_0 \ar[r]^-{\rm Ad} & G_0,
}
\end{equation}
commutes.
\end{enumerate}
In the description above,
${\rm Ad}: G \rightarrow {\rm Aut}(G)$ denotes the
adjoint action of $G$ to itself, namely 
Ad$(g)(x) = g x g^{-1}$.

Some examples of crossed modules include the following:
\begin{itemize}
\item For $G_1$ any group, let $G_0 = {\rm Aut}(G_1)$,
with $d: G_1 \rightarrow {\rm Aut}(G_1)$ the natural inclusion
(meaning $d(g) = {\rm Ad}(g)$)
and $\alpha: {\rm Aut}(G_1) \rightarrow {\rm Aut}(G_1)$ the identity.
\item Let $G_0$ be any group and $G_1$ a normal subgroup of $G_0$,
with $d: G_1 \rightarrow G_0$ inclusion, and $\alpha: G_0 \rightarrow
{\rm Aut}(G_1)$ by conjugation.
\end{itemize}

A crossed module can be encoded in a four-term exact sequence:
\begin{equation}
1 \: \longrightarrow \: {\rm Ker}\, d \: \longrightarrow \:
G_1 \: \stackrel{d}{\longrightarrow} \: G_0 \:
\longrightarrow \: {\rm Coker}\, d \: \longrightarrow \: 1.
\end{equation}
In the case that Ker $d$ is abelian, this is sometimes alternatively
expressed as the extension
\begin{equation}
1 \: \longrightarrow \: B({\rm Ker}\,d) \: \longrightarrow \:
\Gamma \: \longrightarrow \: {\rm Coker}\, d \: \longrightarrow \: 1,
\end{equation}
for $\Gamma$ the 2-group corresonding to the crossed module.

Physically, in this paper, the map $d$ encodes the action of the
noneffectively-acting $BA$, by mapping $A$ to a subset of the
center of the Chern-Simons gauge group, which acts nontrivially.

For more information on crossed modules,
see for example \cite{bhs} for further mathematics background,
or \cite[appendix A]{Lee:2021crt}, \cite[section 2]{Bhardwaj:2021wif} in physics.

\section{Generalities on gauging effectively-acting one-form symmetries}
\label{sect:gauging-bk}

For most of this paper, we have discussed gauging one-form symmetries
in terms of line operators, but it is worth observing that this operation
can also be understood in terms of local actions, which we will briefly
review in
this section.

Suppose in general terms we have a $G$ gauge theory, and we gauge
the action of a one-form symmetry $BK$, where $BK$ acts nontrivially
on the line operators of the theory.  (For example, this is the case if
$K$ is a subset of the center of $G$.)

In general terms, when gauging the $BK$ on a $G$ gauge theory,
\begin{itemize}
\item the path integral sums over $K$ gerbes, and
\item for each $K$ gerbe, the path integral sums over gerbe-twisted
$G$ bundles, defined by transition functions which close on triple overlaps
only up to a cocycle representing the gerbe characteristic class.
\end{itemize}

Consider for example gauging an effectively-acting $B {\mathbb Z}_n$
in an $SU(n)$ gauge theory.
The twisted $SU(n)$ gauge fields above are all the same as
ordinary $SU(n)/{\mathbb Z}_n$ gauge fields, and the gerbe characteristic
classes correspond to (some) characteristic classes of $SU(n)/{\mathbb Z}_n$
bundles.  Let us look at this in more detail:
\begin{enumerate}
\item The transition functions $g_{ij}$ of a twisted bundle no longer close
on triple overlaps, but rather obey
\begin{equation}
g_{ij} g_{jk} g_{ki} \: = \: h_{ijk}
\end{equation}
for a cocycle $h_{ijk}$ representing an element of $H^2(Y, {\mathbb Z}_n)$
corresponding to the gerbe characteristic class, and
\item Across overlaps, the gauge field $A$ obeys
\begin{equation}
A_i \: = \: g_{ij} A_j g_{ij}^{-1} \: + \:
g_{ij}^{-1} dg_{ij} \: - \: I \Lambda_{ij},
\end{equation}
where $I$ is the identity and
$\Lambda_{ij}$ is a locally-defined one-form field, with the property
that if the gerbe were to admit a connection $B$, then on the same overlaps
\begin{equation}
B_i \: = \: B_{j} \: + \: d \Lambda_{ij}.
\end{equation}
\end{enumerate}

Now, this procedure should generate all $G/K$ bundles.
One example of this involves the relation between $SU(2)$ and $SO(3)$ bundles
in three-dimensional theories.  As is well-known,
\begin{equation}
\mbox{Chern-Simons}(SU(2)) / B {\mathbb Z}_2 \: = \:
\mbox{Chern-Simons}(SO(3)),
\end{equation}
for the $B {\mathbb Z}_2$ corresponding to the center one-form symmetry.
Viewed as a $B {\mathbb Z}_2$ quotient of an $SU(2)$ gauge theory, the
path integral
\begin{itemize}
\item sums over ${\mathbb Z}_2$ gerbes, whose characteristic class is
$w \in H^2(M,{\mathbb Z}_2)$, and
\item sums over $w$-twisted $SU(2)$ bundles, meaning that the $SU(2)$
transition functions close on triple overlaps only up to $w$,
and that gauge transformations across patches only have to match up to
a ${\mathbb Z}_2$ shift.
\end{itemize}
Interpreted in terms of $SO(3)$ bundles, the characteristic class
$w \in H^2(M, {\mathbb Z}_2)$ is the second Stiefel-Whitney class of
an $SO(3)$ bundle.  (The other possibly nonzero characteristic class,
the third Stiefel-Whitney class $w_3 \in H^3(M,{\mathbb Z}_2)$,
is determined by $w_2$ as $w_3 = {\rm Sq}^1(w_2)$, 
see section~\ref{sect:ex:su2-bz4}.)
The fact that gauge transformations only respect $SU(2)$ up to ${\mathbb Z}_2$
shifts, and that $SU(2)$ transition functions only close up to $w$, are
indicative of general aspects of $SO(3)$ bundles.

Thus, we see that the $B {\mathbb Z}_2$-gauged $SU(2)$ theory really does
recover all $SO(3)$ bundles, even those with nonzero $w_3$,
as expected.

If we instead gauged a $BA$ action on a $G$ Chern-Simons theory
with a trivially-acting subgroup $BK$,
then, for reasons detailed in \cite{Pantev:2022kpl},
we would recover $G / (A/K)$ gauge theory, with a restriction on
$G / (A/K)$ bundles.  One role of decomposition is to implement that
restriction.

\end{document}